\documentclass[reprint,amsmath,amssymb,eqsecnum,aip]{revtex4-1}

\usepackage{dcolumn}

    \usepackage{bm}
    \usepackage{graphicx}
\newcommand\beq{\begin{equation}}
\newcommand\be{\begin{equation}}
\newcommand\ee{\end{equation}}
\newcommand\eeq{\end{equation}}
\newcommand\beqa{\begin{eqnarray}}
\newcommand\eeqa{\end{eqnarray}}
\newcommand{\nn}{\nonumber\\}
\newcommand\ba{\begin{eqnarray}}
\newcommand\ea{\end{eqnarray}}
\newcommand\ocite{\onlinecite}

\def\bal#1\eal{\begin{align}#1\end{align}}

\usepackage{epic}
\newcommand{\xx}{5}
\newcommand{\yy}{5}
\newcommand{\yyfive}{8}
\newcommand{\xmax}{30}
\newcommand{\xmaxnew}{25}
\newcommand{\ymax}{22}
\newcommand{\ymaxpenta}{35}
\newcommand{\newunit}{.1mm}

\newcommand{\SfourbinA}{\begin{picture}(\xmax,\ymax)(-\xx,\yy)
\setlength{\unitlength}{.1mm}
\put(0,0){\circle{18}}\put(0,0){\circle*{9}}
\put(60,0){\circle*{18}}
\put(60,60){\circle*{18}}
\put(0,60){\circle*{18}}
\put(9,0){\line(1,0){42}}
\put(0,9){\line(0,1){42}}
\put(60,9){\line(0,1){42}}
\put(9,60){\line(1,0){42}}
\end{picture}}

\newcommand{\SfourbinB}{\begin{picture}(\xmax,\ymax)(-\xx,\yy)
\setlength{\unitlength}{.1mm}
\put(0,0){\circle{18}}\put(0,0){\circle*{9}}
\put(60,0){\circle*{18}}
\put(60,60){\circle*{18}}
\put(0,60){\circle*{18}}
\put(9,0){\line(1,0){42}}
\put(7,7){\line(1,1) {46.5}}
\put(0,9){\line(0,1){42}}
\put(60,9){\line(0,1){42}}
\put(9,60){\line(1,0){42}}
\end{picture}}

\newcommand{\SfourbinC}{\begin{picture}(\xmax,\ymax)(-\xx,\yy)
\setlength{\unitlength}{.1mm}
\put(0,0){\circle*{18}}\put(0,0){\circle*{9}}
\put(60,0){\circle{18}}\put(60,0){\circle*{9}}
\put(60,60){\circle*{18}}
\put(0,60){\circle*{18}}
\put(9,0){\line(1,0){42}}
\put(7,7){\line(1,1) {46.5}}
\put(0,9){\line(0,1){42}}
\put(60,9){\line(0,1){42}}
\put(9,60){\line(1,0){42}}
\end{picture}}

\newcommand{\SfourbinD}{\begin{picture}(\xmax,\ymax)(-\xx,\yy)
\setlength{\unitlength}{.1mm}
\put(0,0){\circle{18}}\put(0,0){\circle*{9}}
\put(60,0){\circle*{18}}
\put(60,60){\circle*{18}}
\put(0,60){\circle*{18}}
\put(9,0){\line(1,0){42}}
\put(7,7){\line(1,1) {46.5}}
\put(0,9){\line(0,1){42}}
\put(60,9){\line(0,1){42}}
\put(7,53){\line(1,-1) {46.5}}
\put(9,60){\line(1,0){42}}
\end{picture}}

\newcommand{\SfourbinAA}{\begin{picture}(\xmax,\ymax)(-\xx,\yy)
\setlength{\unitlength}{.1mm}
\put(0,0){\circle{18}}\put(0,0){\circle*{9}}
\put(60,0){\circle{18}}\put(60,0){\circle*{9}}
\put(60,60){\circle*{18}}
\put(0,60){\circle*{18}}
\put(9,0){\line(1,0){42}}
\put(0,9){\line(0,1){42}}
\put(60,9){\line(0,1){42}}
\put(9,60){\line(1,0){42}}
\end{picture}}

\newcommand{\SfourbinAB}{\begin{picture}(\xmax,\ymax)(-\xx,\yy)
\setlength{\unitlength}{.1mm}
\put(0,0){\circle{18}}\put(0,0){\circle*{9}}
\put(60,0){\circle*{18}}\put(60,0){\circle*{9}}
\put(60,60){\circle{18}}\put(60,60){\circle*{9}}
\put(0,60){\circle*{18}}
\put(9,0){\line(1,0){42}}
\put(0,9){\line(0,1){42}}
\put(60,9){\line(0,1){42}}
\put(9,60){\line(1,0){42}}
\end{picture}}

\newcommand{\SfourbinAC}{\begin{picture}(\xmax,\ymax)(-\xx,\yy)
\setlength{\unitlength}{.1mm}
\put(0,0){\circle{18}}\put(0,0){\circle*{9}}
\put(60,0){\circle{18}}\put(60,0){\circle*{9}}
\put(60,60){\circle*{18}}
\put(0,60){\circle*{18}}
\put(9,0){\line(1,0){42}}
\put(7,7){\line(1,1) {46.5}}
\put(0,9){\line(0,1){42}}
\put(60,9){\line(0,1){42}}
\put(9,60){\line(1,0){42}}
\end{picture}}

\newcommand{\SfourbinAD}{\begin{picture}(\xmax,\ymax)(-\xx,\yy)
\setlength{\unitlength}{.1mm}
\put(0,0){\circle{18}}\put(0,0){\circle*{9}}
\put(60,0){\circle*{18}}\put(60,0){\circle*{9}}
\put(60,60){\circle{18}}\put(60,60){\circle*{9}}
\put(0,60){\circle*{18}}
\put(9,0){\line(1,0){42}}
\put(7,7){\line(1,1) {46.5}}
\put(0,9){\line(0,1){42}}
\put(60,9){\line(0,1){42}}
\put(9,60){\line(1,0){42}}
\end{picture}}

\newcommand{\SfourbinAE}{\begin{picture}(\xmax,\ymax)(-\xx,\yy)
\setlength{\unitlength}{.1mm}
\put(0,0){\circle*{18}}\put(0,0){\circle*{9}}
\put(60,0){\circle{18}}\put(60,0){\circle*{9}}
\put(60,60){\circle*{18}}
\put(0,60){\circle{18}}\put(0,60){\circle*{9}}
\put(9,0){\line(1,0){42}}
\put(7,7){\line(1,1) {46.5}}
\put(0,9){\line(0,1){42}}
\put(60,9){\line(0,1){42}}
\put(9,60){\line(1,0){42}}
\end{picture}}

\newcommand{\SfourbinAF}{\begin{picture}(\xmax,\ymax)(-\xx,\yy)
\setlength{\unitlength}{.1mm}
\put(0,0){\circle{18}}\put(0,0){\circle*{9}}
\put(60,0){\circle{18}}\put(60,0){\circle*{9}}
\put(60,60){\circle*{18}}
\put(0,60){\circle*{18}}
\put(9,0){\line(1,0){42}}
\put(7,7){\line(1,1) {46.5}}
\put(0,9){\line(0,1){42}}
\put(60,9){\line(0,1){42}}
\put(7,53){\line(1,-1) {46.5}}
\put(9,60){\line(1,0){42}}
\end{picture}}

\newcommand{\SfiveA}{\begin{picture}(\xmaxnew,\ymaxpenta)(-\xx,\yyfive)
\setlength{\unitlength}{\newunit}
\put(0,0){\circle*{18}}
\put(60,0){\circle*{18}}
\put(60,60){\circle*{18}}
\put(0,60){\circle*{18}}
\put(9,0){\line(1,0){42}}
\put(0,9){\line(0,1){42}}
\put(60,9){\line(0,1){42}}
\put(30,90){\line(1,-1){36}}
\put(30,90){\line(-1,-1){36}}
\put(30,90){\circle*{9}}\put(30,90){\circle{18}}
\end{picture}}

\newcommand{\SfiveB}{\begin{picture}(\xmaxnew,\ymaxpenta)(-\xx,\yyfive)
\setlength{\unitlength}{.1mm}
\put(0,0){\circle*{9}}\put(0,0){\circle{18}}
\put(60,0){\circle*{18}}
\put(60,60){\circle*{18}}
\put(0,60){\circle*{18}}
\put(9,0){\line(1,0){42}}
\put(0,9){\line(0,1){42}}
\put(60,9){\line(0,1){42}}
\put(30,90){\line(1,-1){36}}
\put(60,60){\line(-1,0){60}}
\put(30,90){\line(-1,-1){36}}
\put(30,90){\circle*{18}}
\end{picture}}

\newcommand{\SfiveC}{\begin{picture}(\xmaxnew,\ymaxpenta)(-\xx,\yyfive)
\setlength{\unitlength}{.1mm}
\put(0,0){\circle*{18}}
\put(60,0){\circle*{18}}
\put(60,60){\circle*{18}}
\put(30,90){\circle*{18}}
\put(0,60){\circle*{9}}\put(0,60){\circle{18}}
\put(9,0){\line(1,0){42}}
\put(0,9){\line(0,1){42}}
\put(60,9){\line(0,1){42}}
\put(30,90){\line(1,-1){36}}
\put(60,60){\line(-1,0){60}}
\put(30,90){\line(-1,-1){36}}
\end{picture}}

\newcommand{\SfiveD}{\begin{picture}(\xmaxnew,\ymaxpenta)(-\xx,\yyfive)
\setlength{\unitlength}{.1mm}
\put(0,0){\circle*{18}}
\put(60,0){\circle*{18}}
\put(60,60){\circle*{18}}
\put(30,90){\circle*{9}}\put(30,90){\circle{18}}
\put(0,60){\circle*{18}}
\put(9,0){\line(1,0){42}}
\put(0,9){\line(0,1){42}}
\put(60,9){\line(0,1){42}}
\put(30,90){\line(1,-1){36}}
\put(60,60){\line(-1,0){60}}
\put(30,90){\line(-1,-1){36}}
\end{picture}}

\newcommand{\SfiveE}{\begin{picture}(\xmaxnew,\ymaxpenta)(-\xx,\yyfive)
\setlength{\unitlength}{.1mm}
\put(0,0){\circle*{18}}
\put(60,0){\circle*{18}}
\put(60,60){\circle*{18}}
\put(30,90){\circle*{9}}\put(30,90){\circle{18}}
\put(0,60){\circle*{18}}
\put(7,7){\line(1,1){46.5}}
\put(0,9){\line(0,1){42}}
\put(60,9){\line(0,1){42}}
\put(53,7){\line(-1,1){46.5}}
\put(30,90){\line(1,-1){36}}
\put(30,90){\line(-1,-1){36}}
\end{picture}}

\newcommand{\SfiveF}{\begin{picture}(\xmaxnew,\ymaxpenta)(-\xx,\yyfive)
\setlength{\unitlength}{.1mm}
\put(0,0){\circle*{18}}
\put(60,0){\circle*{18}}
\put(60,60){\circle*{18}}
\put(30,90){\circle*{18}}
\put(0,60){\circle{18}}\put(0,60){\circle*{9}}
\put(7,7){\line(1,1){46.5}}
\put(0,9){\line(0,1){42}}
\put(60,9){\line(0,1){42}}
\put(53,7){\line(-1,1){46.5}}
\put(30,90){\line(1,-1){36}}
\put(30,90){\line(-1,-1){36}}
\end{picture}}

\newcommand{\SfiveG}{\begin{picture}(\xmaxnew,\ymaxpenta)(-\xx,\yyfive)
\setlength{\unitlength}{.1mm}
\put(0,0){\circle*{18}}
\put(60,0){\circle*{18}}
\put(60,60){\circle*{18}}
\put(30,90){\circle*{9}}\put(30,90){\circle{18}}
\put(0,60){\circle*{18}}
\put(7,7){\line(1,1){46.5}}
\put(0,9){\line(0,1){42}}
\put(60,9){\line(0,1){42}}
\put(53,7){\line(-1,1){46.5}}
\put(30,90){\line(1,-1){36}}
\put(60,60){\line(-1,0){60}}
\put(30,90){\line(-1,-1){36}}
\end{picture}}

\newcommand{\SfiveH}{\begin{picture}(\xmaxnew,\ymaxpenta)(-\xx,\yyfive)
\setlength{\unitlength}{.1mm}
\put(0,0){\circle*{18}}
\put(60,0){\circle*{18}}
\put(60,60){\circle*{18}}
\put(30,90){\circle*{18}}
\put(0,60){\circle{18}}\put(0,60){\circle*{9}}
\put(7,7){\line(1,1){46.5}}
\put(0,9){\line(0,1){42}}
\put(60,9){\line(0,1){42}}
\put(53,7){\line(-1,1){46.5}}
\put(30,90){\line(1,-1){36}}
\put(60,60){\line(-1,0){60}}
\put(30,90){\line(-1,-1){36}}
\end{picture}}

\newcommand{\SfiveI}{\begin{picture}(\xmaxnew,\ymaxpenta)(-\xx,\yyfive)
\setlength{\unitlength}{.1mm}
\put(0,0){\circle*{9}}\put(0,0){\circle{18}}
\put(60,0){\circle*{18}}
\put(60,60){\circle*{18}}
\put(0,60){\circle*{18}}
\put(9,0){\line(1,0){42}}
\put(2.8,8.5){\line(1,3){27}}
\put(0,9){\line(0,1){42}}
\put(60,9){\line(0,1){42}}
\put(57.2,8.5){\line(-1,3){27}}
\put(30,90){\line(1,-1){36}}
\put(30,90){\line(-1,-1){36}}
\put(30,90){\circle*{18}}
\end{picture}}

\newcommand{\SfiveJ}{\begin{picture}(\xmaxnew,\ymaxpenta)(-\xx,\yyfive)
\setlength{\unitlength}{.1mm}
\put(0,0){\circle*{18}}
\put(60,0){\circle*{18}}
\put(60,60){\circle*{18}}
\put(0,60){\circle{18}}\put(0,60){\circle*{9}}
\put(9,0){\line(1,0){42}}
\put(2.8,8.5){\line(1,3){27}}
\put(0,9){\line(0,1){42}}
\put(60,9){\line(0,1){42}}
\put(57.2,8.5){\line(-1,3){27}}
\put(30,90){\line(1,-1){36}}
\put(30,90){\line(-1,-1){36}}
\put(30,90){\circle*{18}}
\end{picture}}

\newcommand{\SfiveK}{\begin{picture}(\xmaxnew,\ymaxpenta)(-\xx,\yyfive)
\setlength{\unitlength}{.1mm}
\put(0,0){\circle*{18}}
\put(60,0){\circle*{18}}
\put(60,60){\circle*{18}}
\put(30,90){\circle{18}}\put(30,90){\circle*{9}}
\put(0,60){\circle*{18}}
\put(9,0){\line(1,0){42}}
\put(2.8,8.5){\line(1,3){27}}
\put(0,9){\line(0,1){42}}
\put(60,9){\line(0,1){42}}
\put(57.2,8.5){\line(-1,3){27}}
\put(30,90){\line(1,-1){36}}
\put(30,90){\line(-1,-1){36}}
\end{picture}}

\newcommand{\SfiveL}{\begin{picture}(\xmaxnew,\ymaxpenta)(-\xx,\yyfive)
\setlength{\unitlength}{.1mm}
\put(0,0){\circle{18}}\put(0,0){\circle*{9}}
\put(60,0){\circle*{18}}
\put(60,60){\circle*{18}}
\put(30,90){\circle*{18}}
\put(0,60){\circle*{18}}
\put(9,0){\line(1,0){42}}
\put(7,7){\line(1,1){46.5}}
\put(0,9){\line(0,1){42}}
\put(60,9){\line(0,1){42}}
\put(53,7){\line(-1,1){46.5}}
\put(30,90){\line(1,-1){36}}
\put(30,90){\line(-1,-1){36}}
\end{picture}}

\newcommand{\SfiveM}{\begin{picture}(\xmaxnew,\ymaxpenta)(-\xx,\yyfive)
\setlength{\unitlength}{.1mm}
\put(0,0){\circle*{18}}
\put(60,0){\circle*{18}}
\put(60,60){\circle*{18}}
\put(30,90){\circle*{18}}
\put(0,60){\circle{18}}\put(0,60){\circle*{9}}
\put(9,0){\line(1,0){42}}
\put(7,7){\line(1,1){46.5}}
\put(0,9){\line(0,1){42}}
\put(60,9){\line(0,1){42}}
\put(53,7){\line(-1,1){46.5}}
\put(30,90){\line(1,-1){36}}
\put(30,90){\line(-1,-1){36}}
\end{picture}}

\newcommand{\SfiveN}{\begin{picture}(\xmaxnew,\ymaxpenta)(-\xx,\yyfive)
\setlength{\unitlength}{.1mm}
\put(0,0){\circle*{18}}
\put(60,0){\circle*{18}}
\put(60,60){\circle*{18}}
\put(30,90){\circle{18}}\put(30,90){\circle*{9}}
\put(0,60){\circle*{18}}
\put(9,0){\line(1,0){42}}
\put(7,7){\line(1,1){46.5}}
\put(0,9){\line(0,1){42}}
\put(60,9){\line(0,1){42}}
\put(53,7){\line(-1,1){46.5}}
\put(30,90){\line(1,-1){36}}
\put(30,90){\line(-1,-1){36}}
\end{picture}}

\newcommand{\SfiveO}{\begin{picture}(\xmaxnew,\ymaxpenta)(-\xx,\yyfive)
\setlength{\unitlength}{.1mm}
\put(0,0){\circle*{18}}
\put(60,0){\circle*{18}}
\put(60,60){\circle*{18}}
\put(30,90){\circle*{18}}
\put(0,60){\circle{18}}\put(0,60){\circle*{9}}
\put(9,0){\line(1,0){42}}
\put(2.8,8.5){\line(1,3){27}}
\put(0,9){\line(0,1){42}}
\put(60,9){\line(0,1){42}}
\put(57.2,8.5){\line(-1,3){27}}
\put(30,90){\line(1,-1){36}}
\put(60,60){\line(-1,0){60}}
\put(30,90){\line(-1,-1){36}}
\end{picture}}

\newcommand{\SfiveP}{\begin{picture}(\xmaxnew,\ymaxpenta)(-\xx,\yyfive)
\setlength{\unitlength}{.1mm}
\put(0,0){\circle*{18}}
\put(60,0){\circle*{18}}
\put(60,60){\circle*{18}}
\put(30,90){\circle{18}}\put(30,90){\circle*{9}}
\put(0,60){\circle*{18}}
\put(9,0){\line(1,0){42}}
\put(2.8,8.5){\line(1,3){27}}
\put(0,9){\line(0,1){42}}
\put(60,9){\line(0,1){42}}
\put(57.2,8.5){\line(-1,3){27}}
\put(30,90){\line(1,-1){36}}
\put(60,60){\line(-1,0){60}}
\put(30,90){\line(-1,-1){36}}
\end{picture}}

\newcommand{\SfiveQ}{\begin{picture}(\xmaxnew,\ymaxpenta)(-\xx,\yyfive)
\setlength{\unitlength}{.1mm}
\put(0,0){\circle{18}}\put(0,0){\circle*{9}}
\put(60,0){\circle*{18}}
\put(60,60){\circle*{18}}
\put(30,90){\circle*{18}}
\put(0,60){\circle*{18}}
\put(9,0){\line(1,0){42}}
\put(7,7){\line(1,1){46.5}}
\put(0,9){\line(0,1){42}}
\put(60,9){\line(0,1){42}}
\put(53,7){\line(-1,1){46.5}}
\put(30,90){\line(1,-1){36}}
\put(60,60){\line(-1,0){60}}
\put(30,90){\line(-1,-1){36}}
\end{picture}}

\newcommand{\SfiveR}{\begin{picture}(\xmaxnew,\ymaxpenta)(-\xx,\yyfive)
\setlength{\unitlength}{.1mm}
\put(0,0){\circle*{18}}
\put(60,0){\circle*{18}}
\put(60,60){\circle*{18}}
\put(30,90){\circle*{18}}
\put(0,60){\circle{18}}\put(0,60){\circle*{9}}
\put(9,0){\line(1,0){42}}
\put(7,7){\line(1,1){46.5}}
\put(0,9){\line(0,1){42}}
\put(60,9){\line(0,1){42}}
\put(53,7){\line(-1,1){46.5}}
\put(30,90){\line(1,-1){36}}
\put(60,60){\line(-1,0){60}}
\put(30,90){\line(-1,-1){36}}
\end{picture}}

\newcommand{\SfiveS}{\begin{picture}(\xmaxnew,\ymaxpenta)(-\xx,\yyfive)
\setlength{\unitlength}{.1mm}
\put(0,0){\circle*{18}}
\put(60,0){\circle*{18}}
\put(60,60){\circle*{18}}
\put(30,90){\circle{18}}\put(30,90){\circle*{9}}
\put(0,60){\circle*{18}}
\put(9,0){\line(1,0){42}}
\put(7,7){\line(1,1){46.5}}
\put(0,9){\line(0,1){42}}
\put(60,9){\line(0,1){42}}
\put(53,7){\line(-1,1){46.5}}
\put(30,90){\line(1,-1){36}}
\put(60,60){\line(-1,0){60}}
\put(30,90){\line(-1,-1){36}}
\end{picture}}

\newcommand{\SfiveT}{\begin{picture}(\xmaxnew,\ymaxpenta)(-\xx,\yyfive)
\setlength{\unitlength}{.1mm}
\put(0,0){\circle*{18}}\put(0,0){\circle*{9}}
\put(60,0){\circle*{18}}
\put(60,60){\circle*{18}}
\put(30,90){\circle{18}}\put(30,90){\circle*{9}}
\put(0,60){\circle*{18}}
\put(9,0){\line(1,0){42}}
\put(7,7){\line(1,1){46.5}}
\put(2.8,8.5){\line(1,3){27}}
\put(0,9){\line(0,1){42}}
\put(60,9){\line(0,1){42}}
\put(57.2,8.5){\line(-1,3){27}}
\put(53,7){\line(-1,1){46.5}}
\put(30,90){\line(1,-1){36}}
\put(30,90){\line(-1,-1){36}}
\end{picture}}

\newcommand{\SfiveU}{\begin{picture}(\xmaxnew,\ymaxpenta)(-\xx,\yyfive)
\setlength{\unitlength}{.1mm}
\put(0,0){\circle*{18}}
\put(60,0){\circle*{18}}
\put(60,60){\circle*{18}}
\put(30,90){\circle*{18}}
\put(0,60){\circle{18}}\put(0,60){\circle*{9}}
\put(9,0){\line(1,0){42}}
\put(7,7){\line(1,1){46.5}}
\put(2.8,8.5){\line(1,3){27}}
\put(0,9){\line(0,1){42}}
\put(60,9){\line(0,1){42}}
\put(57.2,8.5){\line(-1,3){27}}
\put(53,7){\line(-1,1){46.5}}
\put(30,90){\line(1,-1){36}}
\put(30,90){\line(-1,-1){36}}
\end{picture}}

\newcommand{\SfiveV}{\begin{picture}(\xmaxnew,\ymaxpenta)(-\xx,\yyfive)
\setlength{\unitlength}{.1mm}
\put(0,0){\circle*{18}}\put(0,0){\circle*{9}}
\put(60,0){\circle*{18}}
\put(60,60){\circle*{18}}
\put(30,90){\circle{18}}\put(30,90){\circle*{9}}
\put(0,60){\circle*{18}}
\put(9,0){\line(1,0){42}}
\put(7,7){\line(1,1){46.5}}
\put(2.8,8.5){\line(1,3){27}}
\put(0,9){\line(0,1){42}}
\put(60,9){\line(0,1){42}}
\put(57.2,8.5){\line(-1,3){27}}
\put(53,7){\line(-1,1){46.5}}
\put(30,90){\line(1,-1){36}}
\put(60,60){\line(-1,0){60}}
\put(30,90){\line(-1,-1){36}}
\end{picture}}

\newcommand{\SfiveAA}{\begin{picture}(\xmaxnew,\ymaxpenta)(-\xx,\yyfive)
\setlength{\unitlength}{.1mm}
\put(0,0){\circle*{18}}
\put(60,0){\circle*{18}}
\put(60,60){\circle*{18}}
\put(30,90){\circle*{9}}\put(30,90){\circle{18}}
\put(0,60){\circle{18}}\put(0,60){\circle*{9}}
\put(9,0){\line(1,0){42}}
\put(0,9){\line(0,1){42}}
\put(60,9){\line(0,1){42}}
\put(30,90){\line(1,-1){36}}
\put(30,90){\line(-1,-1){36}}
\end{picture}}

\newcommand{\SfiveAB}{\begin{picture}(\xmaxnew,\ymaxpenta)(-\xx,\yyfive)
\setlength{\unitlength}{.1mm}
\put(0,0){\circle{18}}\put(0,0){\circle*{9}}
\put(60,0){\circle*{18}}
\put(60,60){\circle*{18}}
\put(30,90){\circle*{9}}\put(30,90){\circle{18}}
\put(0,60){\circle*{18}}
\put(9,0){\line(1,0){42}}
\put(0,9){\line(0,1){42}}
\put(60,9){\line(0,1){42}}
\put(30,90){\line(1,-1){36}}
\put(30,90){\line(-1,-1){36}}
\end{picture}}

\newcommand{\SfiveAC}{\begin{picture}(\xmaxnew,\ymaxpenta)(-\xx,\yyfive)
\setlength{\unitlength}{.1mm}
\put(0,0){\circle*{18}}
\put(60,0){\circle*{18}}
\put(60,60){\circle*{18}}
\put(30,90){\circle*{9}}\put(30,90){\circle{18}}
\put(0,60){\circle{18}}\put(0,60){\circle*{9}}
\put(9,0){\line(1,0){42}}
\put(0,9){\line(0,1){42}}
\put(60,9){\line(0,1){42}}
\put(30,90){\line(1,-1){36}}
\put(60,60){\line(-1,0){60}}
\put(30,90){\line(-1,-1){36}}
\end{picture}}

\newcommand{\SfiveAD}{\begin{picture}(\xmaxnew,\ymaxpenta)(-\xx,\yyfive)
\setlength{\unitlength}{.1mm}
\put(0,0){\circle{18}}\put(0,0){\circle*{9}}
\put(60,0){\circle*{18}}
\put(60,60){\circle*{18}}
\put(30,90){\circle*{9}}\put(30,90){\circle{18}}
\put(0,60){\circle*{18}}
\put(9,0){\line(1,0){42}}
\put(0,9){\line(0,1){42}}
\put(60,9){\line(0,1){42}}
\put(30,90){\line(1,-1){36}}
\put(60,60){\line(-1,0){60}}
\put(30,90){\line(-1,-1){36}}
\end{picture}}

\newcommand{\SfiveAE}{\begin{picture}(\xmaxnew,\ymaxpenta)(-\xx,\yyfive)
\setlength{\unitlength}{.1mm}
\put(0,0){\circle{18}}\put(0,0){\circle*{9}}
\put(60,0){\circle*{18}}
\put(60,60){\circle*{18}}
\put(30,90){\circle*{18}}
\put(0,60){\circle{18}}\put(0,60){\circle*{9}}
\put(9,0){\line(1,0){42}}
\put(0,9){\line(0,1){42}}
\put(60,9){\line(0,1){42}}
\put(30,90){\line(1,-1){36}}
\put(60,60){\line(-1,0){60}}
\put(30,90){\line(-1,-1){36}}
\end{picture}}

\newcommand{\SfiveAF}{\begin{picture}(\xmaxnew,\ymaxpenta)(-\xx,\yyfive)
\setlength{\unitlength}{.1mm}
\put(0,0){\circle*{18}}
\put(60,0){\circle{18}}\put(60,0){\circle*{9}}
\put(60,60){\circle*{18}}
\put(30,90){\circle*{18}}
\put(0,60){\circle{18}}\put(0,60){\circle*{9}}
\put(9,0){\line(1,0){42}}
\put(0,9){\line(0,1){42}}
\put(60,9){\line(0,1){42}}
\put(30,90){\line(1,-1){36}}
\put(60,60){\line(-1,0){60}}
\put(30,90){\line(-1,-1){36}}
\end{picture}}

\newcommand{\SfiveAG}{\begin{picture}(\xmaxnew,\ymaxpenta)(-\xx,\yyfive)
\setlength{\unitlength}{.1mm}
\put(0,0){\circle*{18}}
\put(60,0){\circle*{18}}
\put(60,60){\circle{18}}\put(60,60){\circle*{9}}
\put(30,90){\circle*{18}}
\put(0,60){\circle{18}}\put(0,60){\circle*{9}}
\put(9,0){\line(1,0){42}}
\put(0,9){\line(0,1){42}}
\put(60,9){\line(0,1){42}}
\put(30,90){\line(1,-1){36}}
\put(60,60){\line(-1,0){60}}
\put(30,90){\line(-1,-1){36}}
\end{picture}}

\newcommand{\SfiveAH}{\begin{picture}(\xmaxnew,\ymaxpenta)(-\xx,\yyfive)
\setlength{\unitlength}{.1mm}
\put(0,0){\circle{18}}\put(0,0){\circle*{9}}
\put(60,0){\circle{18}}\put(60,0){\circle*{9}}
\put(60,60){\circle*{18}}
\put(30,90){\circle*{18}}
\put(0,60){\circle*{18}}
\put(9,0){\line(1,0){42}}
\put(0,9){\line(0,1){42}}
\put(60,9){\line(0,1){42}}
\put(30,90){\line(1,-1){36}}
\put(60,60){\line(-1,0){60}}
\put(30,90){\line(-1,-1){36}}
\end{picture}}

\newcommand{\SfiveAI}{\begin{picture}(\xmaxnew,\ymaxpenta)(-\xx,\yyfive)
\setlength{\unitlength}{.1mm}
\put(0,0){\circle*{18}}
\put(60,0){\circle*{18}}
\put(60,60){\circle*{18}}
\put(30,90){\circle*{9}}\put(30,90){\circle{18}}
\put(0,60){\circle{18}}\put(0,60){\circle*{9}}
\put(7,7){\line(1,1){46.5}}
\put(0,9){\line(0,1){42}}
\put(60,9){\line(0,1){42}}
\put(53,7){\line(-1,1){46.5}}
\put(30,90){\line(1,-1){36}}
\put(30,90){\line(-1,-1){36}}
\end{picture}}

\newcommand{\SfiveAJ}{\begin{picture}(\xmaxnew,\ymaxpenta)(-\xx,\yyfive)
\setlength{\unitlength}{.1mm}
\put(0,0){\circle{18}}\put(0,0){\circle*{9}}
\put(60,0){\circle*{18}}
\put(60,60){\circle*{18}}
\put(30,90){\circle*{9}}\put(30,90){\circle{18}}
\put(0,60){\circle*{18}}
\put(7,7){\line(1,1){46.5}}
\put(0,9){\line(0,1){42}}
\put(60,9){\line(0,1){42}}
\put(53,7){\line(-1,1){46.5}}
\put(30,90){\line(1,-1){36}}
\put(30,90){\line(-1,-1){36}}
\end{picture}}

\newcommand{\SfiveAK}{\begin{picture}(\xmaxnew,\ymaxpenta)(-\xx,\yyfive)
\setlength{\unitlength}{.1mm}
\put(0,0){\circle*{18}}
\put(60,0){\circle*{18}}
\put(60,60){\circle{18}}\put(60,60){\circle*{9}}
\put(30,90){\circle*{18}}
\put(0,60){\circle{18}}\put(0,60){\circle*{9}}
\put(7,7){\line(1,1){46.5}}
\put(0,9){\line(0,1){42}}
\put(60,9){\line(0,1){42}}
\put(53,7){\line(-1,1){46.5}}
\put(30,90){\line(1,-1){36}}
\put(30,90){\line(-1,-1){36}}
\end{picture}}

\newcommand{\SfiveAL}{\begin{picture}(\xmaxnew,\ymaxpenta)(-\xx,\yyfive)
\setlength{\unitlength}{.1mm}
\put(0,0){\circle*{18}}
\put(60,0){\circle*{18}}
\put(60,60){\circle*{18}}
\put(30,90){\circle*{9}}\put(30,90){\circle{18}}
\put(0,60){\circle{18}}\put(0,60){\circle*{9}}
\put(7,7){\line(1,1){46.5}}
\put(0,9){\line(0,1){42}}
\put(60,9){\line(0,1){42}}
\put(53,7){\line(-1,1){46.5}}
\put(30,90){\line(1,-1){36}}
\put(60,60){\line(-1,0){60}}
\put(30,90){\line(-1,-1){36}}
\end{picture}}

\newcommand{\SfiveAM}{\begin{picture}(\xmaxnew,\ymaxpenta)(-\xx,\yyfive)
\setlength{\unitlength}{.1mm}
\put(0,0){\circle{18}}\put(0,0){\circle*{9}}
\put(60,0){\circle*{18}}
\put(60,60){\circle*{18}}
\put(30,90){\circle*{9}}\put(30,90){\circle{18}}
\put(0,60){\circle*{18}}
\put(7,7){\line(1,1){46.5}}
\put(0,9){\line(0,1){42}}
\put(60,9){\line(0,1){42}}
\put(53,7){\line(-1,1){46.5}}
\put(30,90){\line(1,-1){36}}
\put(60,60){\line(-1,0){60}}
\put(30,90){\line(-1,-1){36}}
\end{picture}}

\newcommand{\SfiveAN}{\begin{picture}(\xmaxnew,\ymaxpenta)(-\xx,\yyfive)
\setlength{\unitlength}{.1mm}
\put(0,0){\circle*{18}}
\put(60,0){\circle*{18}}
\put(60,60){\circle{18}}\put(60,60){\circle*{9}}
\put(30,90){\circle*{18}}
\put(0,60){\circle{18}}\put(0,60){\circle*{9}}
\put(7,7){\line(1,1){46.5}}
\put(0,9){\line(0,1){42}}
\put(60,9){\line(0,1){42}}
\put(53,7){\line(-1,1){46.5}}
\put(30,90){\line(1,-1){36}}
\put(60,60){\line(-1,0){60}}
\put(30,90){\line(-1,-1){36}}
\end{picture}}

\newcommand{\SfiveAO}{\begin{picture}(\xmaxnew,\ymaxpenta)(-\xx,\yyfive)
\setlength{\unitlength}{.1mm}
\put(0,0){\circle*{18}}
\put(60,0){\circle*{18}}
\put(60,60){\circle*{18}}
\put(30,90){\circle*{9}}\put(30,90){\circle{18}}
\put(0,60){\circle{18}}\put(0,60){\circle*{9}}
\put(9,0){\line(1,0){42}}
\put(2.8,8.5){\line(1,3){27}}
\put(0,9){\line(0,1){42}}
\put(60,9){\line(0,1){42}}
\put(57.2,8.5){\line(-1,3){27}}
\put(30,90){\line(1,-1){36}}
\put(30,90){\line(-1,-1){36}}
\end{picture}}

\newcommand{\SfiveAP}{\begin{picture}(\xmaxnew,\ymaxpenta)(-\xx,\yyfive)
\setlength{\unitlength}{.1mm}
\put(0,0){\circle{18}}\put(0,0){\circle*{9}}
\put(60,0){\circle*{18}}
\put(60,60){\circle*{18}}
\put(30,90){\circle*{9}}\put(30,90){\circle{18}}
\put(0,60){\circle*{18}}
\put(9,0){\line(1,0){42}}
\put(2.8,8.5){\line(1,3){27}}
\put(0,9){\line(0,1){42}}
\put(60,9){\line(0,1){42}}
\put(57.2,8.5){\line(-1,3){27}}
\put(30,90){\line(1,-1){36}}
\put(30,90){\line(-1,-1){36}}
\end{picture}}

\newcommand{\SfiveAQ}{\begin{picture}(\xmaxnew,\ymaxpenta)(-\xx,\yyfive)
\setlength{\unitlength}{.1mm}
\put(0,0){\circle{18}}\put(0,0){\circle*{9}}
\put(60,0){\circle*{18}}
\put(60,60){\circle*{18}}
\put(30,90){\circle*{18}}
\put(0,60){\circle{18}}\put(0,60){\circle*{9}}
\put(9,0){\line(1,0){42}}
\put(2.8,8.5){\line(1,3){27}}
\put(0,9){\line(0,1){42}}
\put(60,9){\line(0,1){42}}
\put(57.2,8.5){\line(-1,3){27}}
\put(30,90){\line(1,-1){36}}
\put(30,90){\line(-1,-1){36}}
\end{picture}}

\newcommand{\SfiveAR}{\begin{picture}(\xmaxnew,\ymaxpenta)(-\xx,\yyfive)
\setlength{\unitlength}{.1mm}
\put(0,0){\circle*{18}}
\put(60,0){\circle{18}}\put(60,0){\circle*{9}}
\put(60,60){\circle*{18}}
\put(30,90){\circle*{18}}
\put(0,60){\circle{18}}\put(0,60){\circle*{9}}
\put(9,0){\line(1,0){42}}
\put(2.8,8.5){\line(1,3){27}}
\put(0,9){\line(0,1){42}}
\put(60,9){\line(0,1){42}}
\put(57.2,8.5){\line(-1,3){27}}
\put(30,90){\line(1,-1){36}}
\put(30,90){\line(-1,-1){36}}
\end{picture}}

\newcommand{\SfiveAS}{\begin{picture}(\xmaxnew,\ymaxpenta)(-\xx,\yyfive)
\setlength{\unitlength}{.1mm}
\put(0,0){\circle*{18}}
\put(60,0){\circle*{18}}
\put(60,60){\circle{18}}\put(60,60){\circle*{9}}
\put(30,90){\circle*{18}}
\put(0,60){\circle{18}}\put(0,60){\circle*{9}}
\put(9,0){\line(1,0){42}}
\put(2.8,8.5){\line(1,3){27}}
\put(0,9){\line(0,1){42}}
\put(60,9){\line(0,1){42}}
\put(57.2,8.5){\line(-1,3){27}}
\put(30,90){\line(1,-1){36}}
\put(30,90){\line(-1,-1){36}}
\end{picture}}

\newcommand{\SfiveAT}{\begin{picture}(\xmaxnew,\ymaxpenta)(-\xx,\yyfive)
\setlength{\unitlength}{.1mm}
\put(0,0){\circle{18}}\put(0,0){\circle*{9}}
\put(60,0){\circle{18}}\put(60,0){\circle*{9}}
\put(60,60){\circle*{18}}
\put(30,90){\circle*{18}}
\put(0,60){\circle*{18}}
\put(9,0){\line(1,0){42}}
\put(2.8,8.5){\line(1,3){27}}
\put(0,9){\line(0,1){42}}
\put(60,9){\line(0,1){42}}
\put(57.2,8.5){\line(-1,3){27}}
\put(30,90){\line(1,-1){36}}
\put(30,90){\line(-1,-1){36}}
\end{picture}}

\newcommand{\SfiveAU}{\begin{picture}(\xmaxnew,\ymaxpenta)(-\xx,\yyfive)
\setlength{\unitlength}{.1mm}
\put(0,0){\circle{18}}\put(0,0){\circle*{9}}
\put(60,0){\circle*{18}}\put(60,0){\circle*{9}}
\put(60,60){\circle*{18}}
\put(30,90){\circle*{18}}
\put(0,60){\circle{18}}\put(0,60){\circle*{9}}
\put(9,0){\line(1,0){42}}
\put(7,7){\line(1,1){46.5}}
\put(0,9){\line(0,1){42}}
\put(60,9){\line(0,1){42}}
\put(53,7){\line(-1,1){46.5}}
\put(30,90){\line(1,-1){36}}
\put(30,90){\line(-1,-1){36}}
\end{picture}}

\newcommand{\SfiveAV}{\begin{picture}(\xmaxnew,\ymaxpenta)(-\xx,\yyfive)
\setlength{\unitlength}{.1mm}
\put(0,0){\circle*{18}}\put(0,0){\circle*{9}}
\put(60,0){\circle*{18}}\put(60,0){\circle*{9}}
\put(60,60){\circle*{18}}
\put(30,90){\circle{18}}\put(30,90){\circle*{9}}
\put(0,60){\circle{18}}\put(0,60){\circle*{9}}
\put(9,0){\line(1,0){42}}
\put(7,7){\line(1,1){46.5}}
\put(0,9){\line(0,1){42}}
\put(60,9){\line(0,1){42}}
\put(53,7){\line(-1,1){46.5}}
\put(30,90){\line(1,-1){36}}
\put(30,90){\line(-1,-1){36}}
\end{picture}}

\newcommand{\SfiveAX}{\begin{picture}(\xmaxnew,\ymaxpenta)(-\xx,\yyfive)
\setlength{\unitlength}{.1mm}
\put(0,0){\circle{18}}\put(0,0){\circle*{9}}
\put(60,0){\circle*{18}}\put(60,0){\circle*{9}}
\put(60,60){\circle*{18}}
\put(30,90){\circle{18}}\put(30,90){\circle*{9}}
\put(0,60){\circle*{18}}\put(0,60){\circle*{9}}
\put(9,0){\line(1,0){42}}
\put(7,7){\line(1,1){46.5}}
\put(0,9){\line(0,1){42}}
\put(60,9){\line(0,1){42}}
\put(53,7){\line(-1,1){46.5}}
\put(30,90){\line(1,-1){36}}
\put(30,90){\line(-1,-1){36}}
\end{picture}}

\newcommand{\SfiveAY}{\begin{picture}(\xmaxnew,\ymaxpenta)(-\xx,\yyfive)
\setlength{\unitlength}{.1mm}
\put(0,0){\circle*{18}}\put(0,0){\circle*{9}}
\put(60,0){\circle*{18}}\put(60,0){\circle*{9}}
\put(60,60){\circle{18}}\put(60,60){\circle*{9}}
\put(30,90){\circle*{18}}\put(30,90){\circle*{9}}
\put(0,60){\circle{18}}\put(0,60){\circle*{9}}
\put(9,0){\line(1,0){42}}
\put(7,7){\line(1,1){46.5}}
\put(0,9){\line(0,1){42}}
\put(60,9){\line(0,1){42}}
\put(53,7){\line(-1,1){46.5}}
\put(30,90){\line(1,-1){36}}
\put(30,90){\line(-1,-1){36}}
\end{picture}}

\newcommand{\SfiveAZ}{\begin{picture}(\xmaxnew,\ymaxpenta)(-\xx,\yyfive)
\setlength{\unitlength}{.1mm}
\put(0,0){\circle{18}}\put(0,0){\circle*{9}}
\put(60,0){\circle{18}}\put(60,0){\circle*{9}}
\put(60,60){\circle*{18}}\put(60,60){\circle*{9}}
\put(30,90){\circle*{18}}\put(30,90){\circle*{9}}
\put(0,60){\circle*{18}}\put(0,60){\circle*{9}}
\put(9,0){\line(1,0){42}}
\put(7,7){\line(1,1){46.5}}
\put(0,9){\line(0,1){42}}
\put(60,9){\line(0,1){42}}
\put(53,7){\line(-1,1){46.5}}
\put(30,90){\line(1,-1){36}}
\put(30,90){\line(-1,-1){36}}
\end{picture}}

\newcommand{\SfiveBA}{\begin{picture}(\xmaxnew,\ymaxpenta)(-\xx,\yyfive)
\setlength{\unitlength}{.1mm}
\put(0,0){\circle*{18}}\put(0,0){\circle*{9}}
\put(60,0){\circle*{18}}\put(60,0){\circle*{9}}
\put(60,60){\circle*{18}}\put(60,60){\circle*{9}}
\put(30,90){\circle{18}}\put(30,90){\circle*{9}}
\put(0,60){\circle{18}}\put(0,60){\circle*{9}}
\put(9,0){\line(1,0){42}}
\put(2.8,8.5){\line(1,3){27}}
\put(0,9){\line(0,1){42}}
\put(60,9){\line(0,1){42}}
\put(57.2,8.5){\line(-1,3){27}}
\put(30,90){\line(1,-1){36}}
\put(60,60){\line(-1,0){60}}
\put(30,90){\line(-1,-1){36}}
\end{picture}}

\newcommand{\SfiveBB}{\begin{picture}(\xmaxnew,\ymaxpenta)(-\xx,\yyfive)
\setlength{\unitlength}{.1mm}
\put(0,0){\circle{18}}\put(0,0){\circle*{9}}
\put(60,0){\circle*{18}}\put(60,0){\circle*{9}}
\put(60,60){\circle*{18}}\put(60,60){\circle*{9}}
\put(30,90){\circle*{18}}\put(30,90){\circle*{9}}
\put(0,60){\circle{18}}\put(0,60){\circle*{9}}
\put(9,0){\line(1,0){42}}
\put(2.8,8.5){\line(1,3){27}}
\put(0,9){\line(0,1){42}}
\put(60,9){\line(0,1){42}}
\put(57.2,8.5){\line(-1,3){27}}
\put(30,90){\line(1,-1){36}}
\put(60,60){\line(-1,0){60}}
\put(30,90){\line(-1,-1){36}}
\end{picture}}

\newcommand{\SfiveBC}{\begin{picture}(\xmaxnew,\ymaxpenta)(-\xx,\yyfive)
\setlength{\unitlength}{.1mm}
\put(0,0){\circle*{18}}\put(0,0){\circle*{9}}
\put(60,0){\circle{18}}\put(60,0){\circle*{9}}
\put(60,60){\circle*{18}}\put(60,60){\circle*{9}}
\put(30,90){\circle*{18}}\put(30,90){\circle*{9}}
\put(0,60){\circle{18}}\put(0,60){\circle*{9}}
\put(9,0){\line(1,0){42}}
\put(2.8,8.5){\line(1,3){27}}
\put(0,9){\line(0,1){42}}
\put(60,9){\line(0,1){42}}
\put(57.2,8.5){\line(-1,3){27}}
\put(30,90){\line(1,-1){36}}
\put(60,60){\line(-1,0){60}}
\put(30,90){\line(-1,-1){36}}
\end{picture}}

\newcommand{\SfiveBD}{\begin{picture}(\xmaxnew,\ymaxpenta)(-\xx,\yyfive)
\setlength{\unitlength}{.1mm}
\put(0,0){\circle{18}}\put(0,0){\circle*{9}}
\put(60,0){\circle*{18}}\put(60,0){\circle*{9}}
\put(60,60){\circle*{18}}
\put(30,90){\circle*{18}}
\put(0,60){\circle{18}}\put(0,60){\circle*{9}}
\put(9,0){\line(1,0){42}}
\put(7,7){\line(1,1){46.5}}
\put(0,9){\line(0,1){42}}
\put(60,9){\line(0,1){42}}
\put(53,7){\line(-1,1){46.5}}
\put(30,90){\line(1,-1){36}}
\put(60,60){\line(-1,0){60}}
\put(30,90){\line(-1,-1){36}}
\end{picture}}

\newcommand{\SfiveBE}{\begin{picture}(\xmaxnew,\ymaxpenta)(-\xx,\yyfive)
\setlength{\unitlength}{.1mm}
\put(0,0){\circle*{18}}\put(0,0){\circle*{9}}
\put(60,0){\circle*{18}}\put(60,0){\circle*{9}}
\put(60,60){\circle*{18}}
\put(30,90){\circle{18}}\put(30,90){\circle*{9}}
\put(0,60){\circle{18}}\put(0,60){\circle*{9}}
\put(9,0){\line(1,0){42}}
\put(7,7){\line(1,1){46.5}}
\put(0,9){\line(0,1){42}}
\put(60,9){\line(0,1){42}}
\put(53,7){\line(-1,1){46.5}}
\put(30,90){\line(1,-1){36}}
\put(60,60){\line(-1,0){60}}
\put(30,90){\line(-1,-1){36}}
\end{picture}}

\newcommand{\SfiveBF}{\begin{picture}(\xmaxnew,\ymaxpenta)(-\xx,\yyfive)
\setlength{\unitlength}{.1mm}
\put(0,0){\circle{18}}\put(0,0){\circle*{9}}
\put(60,0){\circle*{18}}\put(60,0){\circle*{9}}
\put(60,60){\circle*{18}}
\put(30,90){\circle{18}}\put(30,90){\circle*{9}}
\put(0,60){\circle*{18}}\put(0,60){\circle*{9}}
\put(9,0){\line(1,0){42}}
\put(7,7){\line(1,1){46.5}}
\put(0,9){\line(0,1){42}}
\put(60,9){\line(0,1){42}}
\put(53,7){\line(-1,1){46.5}}
\put(30,90){\line(1,-1){36}}
\put(60,60){\line(-1,0){60}}
\put(30,90){\line(-1,-1){36}}
\end{picture}}

\newcommand{\SfiveBG}{\begin{picture}(\xmaxnew,\ymaxpenta)(-\xx,\yyfive)
\setlength{\unitlength}{.1mm}
\put(0,0){\circle*{18}}\put(0,0){\circle*{9}}
\put(60,0){\circle*{18}}\put(60,0){\circle*{9}}
\put(60,60){\circle{18}}\put(60,60){\circle*{9}}
\put(30,90){\circle*{18}}\put(30,90){\circle*{9}}
\put(0,60){\circle{18}}\put(0,60){\circle*{9}}
\put(9,0){\line(1,0){42}}
\put(7,7){\line(1,1){46.5}}
\put(0,9){\line(0,1){42}}
\put(60,9){\line(0,1){42}}
\put(53,7){\line(-1,1){46.5}}
\put(30,90){\line(1,-1){36}}
\put(60,60){\line(-1,0){60}}
\put(30,90){\line(-1,-1){36}}
\end{picture}}

\newcommand{\SfiveBH}{\begin{picture}(\xmaxnew,\ymaxpenta)(-\xx,\yyfive)
\setlength{\unitlength}{.1mm}
\put(0,0){\circle{18}}\put(0,0){\circle*{9}}
\put(60,0){\circle{18}}\put(60,0){\circle*{9}}
\put(60,60){\circle*{18}}\put(60,60){\circle*{9}}
\put(30,90){\circle*{18}}\put(30,90){\circle*{9}}
\put(0,60){\circle*{18}}\put(0,60){\circle*{9}}
\put(9,0){\line(1,0){42}}
\put(7,7){\line(1,1){46.5}}
\put(0,9){\line(0,1){42}}
\put(60,9){\line(0,1){42}}
\put(53,7){\line(-1,1){46.5}}
\put(30,90){\line(1,-1){36}}
\put(60,60){\line(-1,0){60}}
\put(30,90){\line(-1,-1){36}}
\end{picture}}

\newcommand{\SfiveBI}{\begin{picture}(\xmaxnew,\ymaxpenta)(-\xx,\yyfive)
\setlength{\unitlength}{.1mm}
\put(0,0){\circle*{18}}\put(0,0){\circle*{9}}
\put(60,0){\circle*{18}}\put(60,0){\circle*{9}}
\put(60,60){\circle*{18}}\put(60,60){\circle*{9}}
\put(30,90){\circle{18}}\put(30,90){\circle*{9}}
\put(0,60){\circle{18}}\put(0,60){\circle*{9}}
\put(9,0){\line(1,0){42}}
\put(7,7){\line(1,1){46.5}}
\put(2.8,8.5){\line(1,3){27}}
\put(0,9){\line(0,1){42}}
\put(60,9){\line(0,1){42}}
\put(57.2,8.5){\line(-1,3){27}}
\put(53,7){\line(-1,1){46.5}}
\put(30,90){\line(1,-1){36}}
\put(30,90){\line(-1,-1){36}}
\end{picture}}

\newcommand{\SfiveBJ}{\begin{picture}(\xmaxnew,\ymaxpenta)(-\xx,\yyfive)
\setlength{\unitlength}{.1mm}
\put(0,0){\circle{18}}\put(0,0){\circle*{9}}
\put(60,0){\circle*{18}}\put(60,0){\circle*{9}}
\put(60,60){\circle*{18}}\put(60,60){\circle*{9}}
\put(30,90){\circle{18}}\put(30,90){\circle*{9}}
\put(0,60){\circle*{18}}\put(0,60){\circle*{9}}
\put(9,0){\line(1,0){42}}
\put(7,7){\line(1,1){46.5}}
\put(2.8,8.5){\line(1,3){27}}
\put(0,9){\line(0,1){42}}
\put(60,9){\line(0,1){42}}
\put(57.2,8.5){\line(-1,3){27}}
\put(53,7){\line(-1,1){46.5}}
\put(30,90){\line(1,-1){36}}
\put(30,90){\line(-1,-1){36}}
\end{picture}}

\newcommand{\SfiveBK}{\begin{picture}(\xmaxnew,\ymaxpenta)(-\xx,\yyfive)
\setlength{\unitlength}{.1mm}
\put(0,0){\circle*{18}}\put(0,0){\circle*{9}}
\put(60,0){\circle*{18}}\put(60,0){\circle*{9}}
\put(60,60){\circle{18}}\put(60,60){\circle*{9}}
\put(30,90){\circle*{18}}\put(30,90){\circle*{9}}
\put(0,60){\circle{18}}\put(0,60){\circle*{9}}
\put(9,0){\line(1,0){42}}
\put(7,7){\line(1,1){46.5}}
\put(2.8,8.5){\line(1,3){27}}
\put(0,9){\line(0,1){42}}
\put(60,9){\line(0,1){42}}
\put(57.2,8.5){\line(-1,3){27}}
\put(53,7){\line(-1,1){46.5}}
\put(30,90){\line(1,-1){36}}
\put(30,90){\line(-1,-1){36}}
\end{picture}}

\newcommand{\SfiveBL}{\begin{picture}(\xmaxnew,\ymaxpenta)(-\xx,\yyfive)
\setlength{\unitlength}{.1mm}
\put(0,0){\circle*{18}}\put(0,0){\circle*{9}}
\put(60,0){\circle*{18}}\put(60,0){\circle*{9}}
\put(60,60){\circle*{18}}\put(60,60){\circle*{9}}
\put(30,90){\circle{18}}\put(30,90){\circle*{9}}
\put(0,60){\circle{18}}\put(0,60){\circle*{9}}
\put(9,0){\line(1,0){42}}
\put(7,7){\line(1,1){46.5}}
\put(2.8,8.5){\line(1,3){27}}
\put(0,9){\line(0,1){42}}
\put(60,9){\line(0,1){42}}
\put(57.2,8.5){\line(-1,3){27}}
\put(53,7){\line(-1,1){46.5}}
\put(30,90){\line(1,-1){36}}
\put(60,60){\line(-1,0){60}}
\put(30,90){\line(-1,-1){36}}
\end{picture}}

\begin{document}

\title{Virial coefficients, equation of state, and demixing of  binary asymmetric nonadditive  hard-disk mixtures}

\author{Giacomo Fiumara}
\email{giacomo.fiumara@unime.it}
\affiliation{
Department of Mathematics and Computer Science, Physics and Earth Sciences, University of Messina,
Viale F. Stagno D'Alcontres 31, I-98166 Messina, Italy}
\author{Franz Saija}
\email{franz.saija@cnr.it}
\affiliation{CNR-IPCF, Viale F. Stagno d'Alcontres, 37-98158 Messina, Italy}
\author{Giuseppe Pellicane}
\email{pellicane@ukzn.ac.za}
\affiliation{School of Chemistry and Physics, University of Kwazulu-Natal, Scottsville 3209 and National Institute for Theoretical Physics (NITheP), KZN Node,
Pietermaritzburg, South Africa}
\author{Mariano L\'{o}pez de Haro}
\email{malopez@unam.mx}
\homepage{http://xml.cie.unam.mx/xml/tc/ft/mlh/}  \altaffiliation[on sabbatical leave from ]{Instituto de Energ\'{\i}as Renovables, Universidad Nacional Aut\'onoma de M\'exico (U.N.A.M.),
Temixco, Morelos 62580, M{e}xico}
\author{Andr\'es Santos}
\email{andres@unex.es}
\homepage{http://www.unex.es/eweb/fisteor/andres/}
\author{Santos B. Yuste}
\email{santos@unex.es}
\homepage{http://www.unex.es/eweb/fisteor/santos/}
\affiliation{Departamento de F\'{\i}sica  and Instituto de Computaci\'on Cient\'{\i}fica Avanzada (ICCAEx), Universidad de
Extremadura, Badajoz, E-06006, Spain}
\date{\today}
\begin{abstract}
Values of the fifth virial coefficient, compressibility factors, and fluid-fluid coexistence curves of binary asymmetric nonadditive  mixtures of hard disks are reported. The former correspond to a wide range of size ratios and positive nonadditivities and have been obtained through a standard Monte Carlo method for the computation of the corresponding cluster integrals. The compressibility factors as functions of density, derived from canonical  Monte Carlo simulations, have been obtained for two values of the size ratio ($q=0.4$ and $q=0.5$), a value of the nonadditivity parameter ($\Delta=0.3$), and five values of the mole fraction of the species with the biggest diameter ($x_1=0.1$, $0.3$, $0.5$, $0.7$, and $0.9$). Some points of the coexistence line relative to the fluid-fluid phase transition for the same values of the size ratios and nonadditivity parameter have been obtained from Gibbs Ensemble Monte Carlo simulations. A comparison is made between the numerical results and those that follow from some theoretical equations of state.
\end{abstract}

\date{\today}


\maketitle
\section{Introduction}
\label{intro}

Binary mixtures of hard disks are systems comprising two kinds of disks of diameters $\sigma_1$ and $\sigma_2$, respectively, which interact as follows. They have no interactions at separations larger than a given distance and experience infinite repulsion if their separation is less than that distance. For pairs of the same species, this closest separation is $\sigma_{ii}=\sigma_i$ ($i=1,2$). If the mixture is nonadditive, the closest distance of approach of disks of different species is not the arithmetic mean of $\sigma_{1}$  and $\sigma_{2}$ but rather $\sigma_{12} = \frac{1}{2}(\sigma_{1} + \sigma_{2})(1 + \Delta)$, where the dimensionless parameter $\Delta$ accounts for deviations of the inter-species interactions from additivity and may be either positive or negative. In the former case, which is the one this paper is concerned with, homocoordination is favored and leads to fluid-fluid phase separation in which one phase is rich in disks of species 1 and the other phase is rich in disks of species 2. Henceforth, without loss of generality, we will assume $\sigma_2\leq \sigma_1$.

Although interest in nonadditive hard-disk (NAHD) mixtures dates back to the 1970s and they are both versatile and relatively simple, comparatively speaking there have been much less studies of these systems than the ones devoted to nonadditive three-dimensional hard-sphere mixtures (for a rather complete but non exhaustive list, the reader is referred to Refs.\ \onlinecite{D77,D79,D80,TB78,SS83,MS85,BM88,BM89,BM90,ESH90,FZM91,MC94,N96,ISN97,N97,SFG98,AEH99,HY00,N00,SG02,CRM03,FK03,SHY05,B05,HLL06,BOBZD08,G09,MO10,SHY10,S11,SSYH12,FPPS14,GC16}). Despite this long period, there are still many important issues related to NAHD mixtures that need to be investigated. For instance, there are a few proposals for the equation of state (EoS) whose merits have not been fully evaluated. Also, apart from the second and third virial coefficients (which are known analytically), higher order virial coefficients require the numerical evaluation of cluster integrals among groups of disks, a task which is in general difficult.

In 2005, three of us\cite{SHY05} introduced an approximate EoS for nonadditive hard-core systems in $d$ dimensions and, taking $d=2$, compared the results obtained for the corresponding compressibility factor with simulation data. Later, a unified framework for some of the most important theories (including some generalizations) of the EoS of $d$-dimensional nonadditive hard-core mixtures was presented.\cite{SHY10}
In 2011, another of us\cite{S11} computed the fourth virial coefficient of \emph{symmetric} NAHD mixtures ($\sigma_2=\sigma_1$) over a wide range of nonadditivity and   compared the fluid-fluid coexistence curve derived from two EoSs built using the new virial coefficients with some simulation results. More recently, four of us\cite{SSYH12} extended the previous study to \emph{asymmetric} NAHD mixtures, ({i.e.}, mixtures such that the size ratio $q = \sigma_{2} / \sigma_{1}$ is different from unity), while three of us\cite{FPPS14} reported values of the fifth virial coefficient and studied the EoS and the equilibrium behavior of a symmetric NAHD mixture.

One of the major aims of this paper is to present the results of computations of the fifth virial coefficient of  binary asymmetric NAHD mixtures. We will explore a range of positive values of the nonadditivity parameter ($\Delta=0.05$, $0.1$, $0.15$, $0.2$, $0.4$, $0.5$, and $0.6$) and  size ratios ($q=0.1$, $0.2$, $0.3$, $0.4$, $0.5$, $0.6667$, $0.8$, $0.85$, $0.9$, and $0.95$). These results complement the ones already published for symmetric mixtures\cite{S11} and will afterwards be used to assess the merits and limitations of some theoretical EoSs.
The compressibility factor predicted by those approximations,  including the rescaled virial expansion that one may construct from the knowledge of the first five virial coefficients, will also be tested against our Monte Carlo (MC) simulation values for a nonadditivity parameter $\Delta=0.3$ and size ratios $q=0.4$ and $q=0.5$. Additionally, we will  address the question of demixing for the same mixtures. As far as we know, apart from the work of Gu\'aqueta,\cite{G09} who used a combination of MC techniques to determine the location of the critical consolute point of asymmetric NAHD mixtures for a wide range of size ratios and values of the positive nonadditivity, simulation results of fluid-fluid separation in NAHD mixtures have only been reported for the symmetric case using either the Gibbs ensemble MC (GEMC) method,\cite{N96,ISN97,N97,N00,FPPS14} the constant volume NVT ensemble MC method,\cite{SG02} the cluster algorithms that allow the consideration of systems with a very large number of particles,\cite{B05,GC16} or the semigrand canonical ensemble MC method.\cite{MO10} Here we will report  GEMC results for the corresponding fluid-fluid coexistence curves of binary asymmetric NAHD mixtures and compare them to the theoretical predictions stemming out of the theoretical EoSs. Once more, our results will complement those of the symmetric case reported in Ref.\ \onlinecite{FPPS14}.

The paper is organized as follows. In Sec.\ \ref{sec2} we recall the well known analytical results for the second and third virial coefficients of a NAHD mixture, and provide the graphical representation of the (partial) composition-independent fourth  and fifth virial coefficients. For these, we provide numerical (MC) results for a wide range of size ratios and positive nonadditivities. We also recall in this section two approximate EoSs proposed by Hamad and by three of us, respectively, and, from the knowledge of the new fifth virial coefficients, we also construct another approximate compressibility factor corresponding to a rescaled virial expansion truncated to fifth order. Section \ref{sec3} contains the assessment of the performance of the three previous theoretical EoSs in three respects. On the one hand, the predictions of the partial fourth and fifth virial coefficients of the first two EoSs are compared with the MC data. On the other hand, a comparison between the results of the MC simulations and those derived from the theoretical expressions for the compressibility factors as functions of density and composition is also performed. Finally, the fluid-fluid coexistence curves that follow from the theoretical EoSs are compared to the results of GEMC simulations. The paper is closed in Sec.\ \ref{sec4} with some concluding remarks.

\section{Thermodynamic properties}
\label{sec2}
\subsection{Virial coefficients}
\label{virial}
The virial expansion for a general mixture can be written as
\begin{equation}
\beta p = \rho + B \rho^2 + C \rho^3 + D \rho^4 + E \rho^5 +\cdots,
\label{s:bpr}
\end{equation}
where $p$ is the pressure, $\beta=1/k_BT$ is the inverse temperature (in units of the Boltzmann constant $k_B$), $\rho = \sum_i\rho_i$ is the total number density ($\rho_i$ being the partial number  density of species $i$) and the virial coefficients $B, C, D, E, \ldots$ depend on the relative concentration of the different species and (in the case of a hard-core fluid) on the particle-particle closest distances. Those coefficients have a polynomial dependence on the mole fractions $x_i=\rho_i/\rho$, so they can be expressed in terms of composition-independent virial coefficients as
\begin{subequations}
\beq
B=\sum_{i,j} x_i x_j B_{ij},
\eeq
\beq
C=\sum_{i,j,k} x_i x_j x_kC_{ijk},
\eeq
\beq
\label{ss:ddix}
D=\sum_{i,j,k,\ell} x_i x_j x_k x_\ell D_{ijk\ell},
\eeq
\beq
\label{ss:eeix}
E=\sum_{i,j,k,\ell,m} x_i x_j x_k x_\ell x_mE_{ijk\ell m}.
\eeq
\end{subequations}
In particular, in the case of a binary mixture,
\begin{subequations}
\be
B=x_{1}^2B_{11}+2x_{1} x_2B_{12}+x_2^2B_{22},
\ee
\be
C=x_{1}^3C_{111}+3x_{1}^2x_2C_{112}+3x_{1}x_2^2C_{122}+x_2^3C_{222},
\ee
\bal
D =& x_1^4D_{1111} + 4x_1^3x_2D_{1112} + 6x_1^2x_2^2D_{1122} \nn
    &+4x_1x_2^3D_{1222} +x_2^4 D_{2222},
\label{s:ddix}
\eal
\bal
E =& x_1^5E_{11111} + 5x_1^4x_2E_{11112} + 10x_1^3x_2^2E_{11122} \nn
    &+10x_1^2x_2^3E_{11222}+5x_1x_2^4 E_{12222}+ x_2^5E_{22222}.
\label{s:eeix}
\eal
\end{subequations}

In the special case of hard-disk mixtures (whether additive or not), the composition-independent virial coefficients $B_{ij}$ and $C_{ijk}$ are exact and well known for an arbitrary number of components (see, for instance, Refs.\ \ocite{AEH99,SHY05,S16}). They read
\begin{subequations}
\begin{equation}
B_{ij}=\frac{\pi}{2}\sigma_{ij}^2,
\label{n1x}
\end{equation}
\bal
C_{ijk}=&\frac{\pi}{6}\left[\sigma_{ij}^2 \mathcal{A}_{\sigma_{ik},\sigma_{jk}}(\sigma_{ij})+\sigma_{ik}^2 \mathcal{A}_{\sigma_{ij},\sigma_{jk}}(\sigma_{ik})\right.\nn
&\left.+\sigma_{jk}^2 \mathcal{A}_{\sigma_{ij},\sigma_{ik}}(\sigma_{jk})\right],
\eal
\end{subequations}
where $\mathcal{A}_{a,b}(r)$ is the intersection area of two circles of radii $a$ and $b$ whose centers are separated by a distance $r$. Its explicit expression is
\beq
\mathcal{A}_{a,b}(r)=\begin{cases}
\pi\min(a^2,b^2),& 0\leq r\leq |a-b|,\\
\bar{\mathcal{A}}_{a,b}(r),&|a-b|\leq r\leq a+b,\\
0,&r\geq a+b,
\end{cases}
\label{vv_ab}
\eeq
where
\bal
\bar{\mathcal{A}}_{a,b}(r)=&a^2\cos^{-1}\frac{r^2+a^2-b^2}{2ar}
+b^2\cos^{-1}\frac{r^2+b^2-a^2}{2br}\nn
&-\frac{1}{2}\sqrt{2r^2(a^2+b^2)-(b^2-a^2)^2-r^4}.
\label{27}
\eal
In particular, $\mathcal{A}_{a,a}(a)=\frac{\pi}{8}a^2 b_3$,
$\mathcal{A}_{a,a}(r)=r^2 \mathcal{G}(a/r)$, and $\mathcal{A}_{a,b}(b)=a^2 \mathcal{H}(b/a)$, where $b_3=\frac{16}{3}-\frac{4\sqrt{3}}{\pi}\simeq 3.128\,02$,
\begin{subequations}
\beq
\mathcal{G}(x)=\left(
 2x^2\cos^{-1} \frac{1}{2x}-\sqrt{x^2-\frac{1}{4}}\right)\Theta\left(x-\frac{1}{2}\right),
\label{H.3}
\ee
\bal
\mathcal{H}(x)=&{\pi}x^2-\left[
\left(2x^2-1\right)\cos^{-1} \frac{1}{2x}+\sqrt{x^2-\frac{1}{4}}\right]\nn
&\times\Theta\left(x-\frac{1}{2}\right),
\label{H.4}
\eal
\end{subequations}
$\Theta(x)$ being the Heaviside step function. In Eq.\ \eqref{H.4} use has been made of the mathematical property
$2\cos^{-1}x+\cos^{-1}(1-2x^2)=\pi$.
Thus, in the binary case, one has
\begin{subequations}
\be
C_{111}=\frac{\pi^2}{16}b_3\sigma_{1}^4,\quad C_{222}=\frac{\pi^2}{16}b_3\sigma_{2}^4,
\label{H.9a}
\ee
\be
C_{112}=\frac{\pi^2}{16}b_3\sigma_{1}^4 \mathcal{F}\left(\frac{\sigma_{12}}{\sigma_{1}}\right),\quad
C_{122}=\frac{\pi^2}{16}b_3\sigma_{2}^4 \mathcal{F}\left(\frac{\sigma_{12}}{\sigma_{2}}\right),
\label{H.9b}
\ee
\end{subequations}
where  the function $\mathcal{F}(x)$ is given by
\be
\mathcal{F}(x)= \frac{8}{3\pi b_3}\left[\mathcal{G}(x)+{2}x^2\mathcal{H}(x)\right].
\label{H.13}
\ee

\begin{table}[t]
\caption{Diagrams contributing to the composition-independent virial coefficients $D_{ijk\ell}$. Filled circles and  points enclosed  by open circles denote species $1$ and $2$, respectively.\label{table_Dijkl}}
\begin{ruledtabular}
\begin{tabular}{ll}
$D_{ijk\ell}$ & $\qquad\qquad\qquad\qquad$Diagrams \\
\hline
$D_{1112}$&$=-\displaystyle{\frac{1}{8}}\Bigg(3\SfourbinA+3 \SfourbinB+3\SfourbinC+\SfourbinD\Bigg)$
\\
\noalign{\smallskip}
$D_{1122}$&$=-\displaystyle{\frac{1}{8}}\Bigg(2\SfourbinAA+\SfourbinAB+4\SfourbinAC+\SfourbinAD$\\
&$\qquad\quad+\SfourbinAE+\SfourbinAF\Bigg)$\\
\end{tabular}
\end{ruledtabular}
\end{table}

\begin{table*}[t]
\caption{Diagrams contributing to $E_{ijk\ell m}$.  Filled circles and  points enclosed by open circles denote species $1$ and $2$, respectively. \label{table_Eijklm}}
\begin{ruledtabular}
\begin{tabular}{ll}
$E_{ijk\ell m}$ &$\qquad \qquad\qquad\qquad\qquad\qquad\qquad \qquad\qquad\qquad\qquad\qquad$Diagrams \\
\hline
$E_{11112}$&
$=-\displaystyle{\frac{1}{30}}\Bigg(12\SfiveA+12\SfiveD+24\SfiveC+24\SfiveB+6\SfiveE+4\SfiveF+6\SfiveG
+4\SfiveH+12\SfiveK+24\SfiveJ$\\
&$\qquad\quad+24\SfiveI +6\SfiveN+12\SfiveM +12\SfiveL+3\SfiveP+12\SfiveO+6\SfiveS+12\SfiveR  +12\SfiveQ +6\SfiveT$\\
&$\qquad\quad+4\SfiveU+\SfiveV\Bigg)$
\\
\noalign{\smallskip}
$E_{11122}$&
$=-\displaystyle{\frac{1}{30}}\Bigg(6\SfiveAA  +6\SfiveAB+12\SfiveAC+12\SfiveAD+12\SfiveAE+12\SfiveAF+6\SfiveAG
+6\SfiveAH+6\SfiveAI+3\SfiveAJ$ \\
&$\qquad\qquad+\SfiveAK+6\SfiveAL+3\SfiveAM  +\SfiveAN+12\SfiveAO+12\SfiveAP+12\SfiveAQ+12\SfiveAR+6\SfiveAS  +6\SfiveAT$\\
&$\qquad\quad+6\SfiveAV+6\SfiveAX+12\SfiveAU+3\SfiveAY+3\SfiveAZ  +6\SfiveBA+6\SfiveBB+3\SfiveBC+6\SfiveBE+6\SfiveBF$\\
&$\qquad\quad+12\SfiveBD  +3\SfiveBG+3\SfiveBH+6\SfiveBI+3\SfiveBJ+\SfiveBK+\SfiveBL\Bigg)$\\
\end{tabular}
\end{ruledtabular}
\end{table*}

The fourth-order terms $D_{1111}$ and $D_{2222}$ in Eq.\ \eqref{s:ddix}  can be calculated analytically through the expression of the fourth  virial coefficient for a
monodisperse fluid of particles with diameter $\sigma_{1}$ or
$\sigma_{2}$, respectively, {i.e.},
\be
D_{1111}=\frac{\pi^3}{64}b_4\sigma_1^6,\quad
D_{2222}=\frac{\pi^3}{64}b_4\sigma_2^6,
\ee
where $b_4=8 (2 + 10/\pi^2 - 9 \sqrt{3}/2 \pi)\simeq 4.257\,85$.  Similarly, the terms $E_{11111}$ and $E_{22222}$ in Eq.\ \eqref{s:eeix} are related to the fifth virial coefficient for a
monodisperse fluid of particles with diameter $\sigma_{1}$ or
$\sigma_{2}$, respectively, namely,
\be
E_{11111}=\frac{\pi^4}{256}b_5\sigma_1^8,\quad
E_{22222}=\frac{\pi^4}{256}b_5\sigma_2^8,
\ee
where, although there is no analytical expression for $b_5$, its numerical value is known to be  $b_5\simeq 5.336\,894\,3$.\cite{LKM05} On the other hand, the remaining coefficients $D_{ijk\ell}$ and $E_{ijk\ell m}$ need to be obtained numerically as functions of the size diameters. The diagrams representing the cluster integrals contributing to $D_{1112}$ and $D_{1122}$ are shown in Table \ref{table_Dijkl}, while those corresponding to $E_{11112}$ and $E_{11122}$ are shown in Table \ref{table_Eijklm}.\cite{S16}
The filled circles and the points enclosed by open circles  in each graph appearing in Tables \ref{table_Dijkl} and \ref{table_Eijklm}
identify particles belonging to species $1$ and $2$, respectively. Space integration is carried out over all the vertices of the graph, each bond contributing a factor to the integrand in the form of a Mayer step function. Of course, the coefficients $D_{1222}$, $E_{12222}$ and $E_{11222}$ are obtained from those in Tables \ref{table_Dijkl} and \ref{table_Eijklm} by exchanging $1\leftrightarrow 2$.

\subsection{Theoretical equations of state and corresponding free energies}
\label{EoS}

In this subsection we will recall two theoretical EoSs for NAHD mixtures whose performance with respect to numerical values of the fourth virial coefficients and compressibility factor was proved to be reasonably good in Ref.\ \ocite{SSYH12}. These are the one proposed by Hamad\cite{H94,H96,H96b,HY00} (which we will label with a superscript H) and one proposed by three of us in 2005\cite{SHY05} (labeled with the superscript SHY).

In the former case,  the compressibility factor $Z\equiv{\beta p}/{\rho}$ of the NAHD mixture (with an arbitrary number of components) is given by
\beq
Z^{\text{H}}(\rho)=1+\frac{\pi}{4\xi}\sum_{i,j}x_i
x_j\frac{\sigma_{ij}^2}{X_{ij}}
\left[Z^{\text{pure}}\left(\eta
X_{ij}\right)-1\right],
\label{2.7}
\eeq
where $\eta\equiv \rho \xi $ is the packing fraction of the mixture (with
$\xi \equiv \frac{\pi}{4}\sum_i x_i\sigma_i^2$), $Z^{\text{pure}}(y)$ is the compressibility factor of a one-component hard-disk fluid at
the packing fraction $y$, and
\beq
X_{ij}=\frac{\pi}{2b_3\xi}{\sum_{k}x_kc_{k;ij}}.
\label{2.5}
\eeq
In Eq.\ \eqref{2.5} the coefficients $c_{k;ij}$  are given by
\beq
c_{k,ij}=\frac{4}{\pi}\mathcal{A}_{\sigma_{ik},\sigma_{kj}}(\sigma_{ij}).
\eeq
In particular,
\begin{subequations}
\be
c_{1;11}=\frac{b_3}{2}\sigma_{1}^2,
\label{H.2}
\ee
\be
c_{2;11}=\frac{4}{\pi}\sigma_{1}^2
\mathcal{G}\left(\frac{\sigma_{12}}{\sigma_{1}}\right),\quad
c_{1;12}=\frac{4}{\pi}\sigma_{1}^2
\mathcal{H}\left(\frac{\sigma_{12}}{\sigma_{1}}\right),
\label{H.2b}
\ee\end{subequations}
with the functions $\mathcal{G}(x)$ and $\mathcal{H}(x)$ given by Eqs.\ \eqref{H.3} and \eqref{H.4}, respectively. Other combinations of indices follow from the exchange $1\leftrightarrow 2$ in the above results.

By construction, Eq.\ \eqref{2.7}  yields the correct (exact) second and third virial coefficients of the mixture. However, it gives approximate values for the higher ones. In particular, it leads to the following (approximate) explicit expressions for the  fourth and fifth virial coefficients,\cite{SHY10}
\begin{subequations}
\label{virH}
\beq
\label{vir4H}
D^{\text{H}}=b_4 \frac{\pi}{4}\left(\frac{\pi}{2b_3}\right)^{2}\sum_{i,j}x_i
x_j\sigma_{ij}^2 \left(\sum_kx_kc_{k;ij}\right)^{2},
\eeq
\beq
\label{vir5H}
E^{\text{H}}=b_5 \frac{\pi}{4}\left(\frac{\pi}{2b_3}\right)^{3}\sum_{i,j}x_i
x_j\sigma_{ij}^2 \left(\sum_kx_kc_{k;ij}\right)^{3}.
\eeq
\end{subequations}
The corresponding composition-independent coefficients $D_{ijk\ell}^{\text{H}}$ and $E_{ijk\ell m}^{\text{H}}$ can easily be identified from Eqs.\ \eqref{ss:ddix} and \eqref{ss:eeix}, respectively.

One can derive from Eq.\ \eqref{2.7} the free energy per particle of the mixture  as $ a^{\text{H}}(\rho)=  a_{\text{id}}(\rho)+  a^{\text{H}}_{\text{ex}}(\rho)$, where
\beq
\label{aid}
a_{\text{id}}(\rho)= k_BT\left[ -1+\sum_{i}x_{i}\ln \left(
x_i\rho\lambda_{i}^2\right)\right]
\eeq
is the ideal Helmholtz free energy per particle ($\lambda_{i}$ being the de Broglie wavelength of particles of species $i$) and
\beq
a^{\text{H}}_{\text{ex}}(\rho)= \frac{\pi}{4\xi}\sum_{i,j}\frac{x_ix_j\sigma_{ij}^2}{X_{ij}} a_{\text{ex}}^{\text{pure}}\left(\eta
X_{ij}\right),
\label{2.8}
\eeq
$a_{\text{ex}}^{\text{pure}}(y)$ being the excess Helmholtz free energy per particle of the one-component hard-disk fluid at the packing fraction $y$.

Next, we describe the SHY approach. The associated  compressibility factor reads
\bal
Z^{\text{SHY}}(\rho)=&1+\frac{b_3
B^* -2 C^*}{b_3-2}\frac{\eta}{1-\eta}\nn
& +
\frac{C^*-
B^* }{b_3-2}\left[Z^{\text{pure}}(\eta)-1\right],
\label{new2}
\eal
where we have called $B^*\equiv B/\xi$ and $C^*\equiv C/\xi^2$.
Again, the exact second and third virial coefficients of the mixture are retained. As for the approximate expressions for the fourth and fifth virial coefficients, they are
\begin{subequations}
\label{virSHY}
\beq
D^{\text{SHY}}=\xi\left(\frac{b_4-2}{b_3-2}{C}-\frac{b_4-b_3}{b_3-2}\xi {B}\right),
 \label{4D.1}
\eeq
\beq
{E^{\text{SHY}}}=\xi^{2}\left(\frac{b_5-2}{b_3-2}{C}-\frac{b_5-b_3}{b_3-2} \xi{B}\right).
 \label{5D.1}
\eeq
\end{subequations}
As before, the composition-independent coefficients $D_{ijk\ell}^{\text{SHY}}$ and $E_{ijk\ell m}^{\text{SHY}}$ can easily be extracted.

The corresponding excess
Helmholtz free energy per particle  is given by
\beq
a^{\text{SHY}}_{\text{ex}}(\rho)= -\frac{b_3 B^*-2
C^*}{b_3-2} \ln(1-\eta)+
\frac{C^*- B^*}{b_3-2} a_{\text{ex}}^{\text{pure}}(\eta).
\label{FEN-SYH}
\eeq

In Ref.\ \ocite{SSYH12} we assessed the merits of the above approximations with respect to the partial fourth virial coefficients. We will do the same with the fifth virial coefficients in Sec.\ \ref{sec3A}. But before we close this section, we will also make use of the numerical knowledge of the fourth and fifth virial coefficients to construct the corresponding rescaled virial expansion (RVE) (as proposed by Baus and Colot\cite{BC87,BXHB88}) truncated to the fifth order. This leads to the following (approximate) EoS for an asymmetric NAHD mixture
\begin{equation}
Z^{\text{RVE}}(\rho)=\frac{1 + c_1\eta + c_2\eta^{2} + c_3\eta^{3}+c_4\eta^{4}}{(1 - \eta)^{2}},
\label{rve}
\end{equation}
where the coefficients $c_1$--$c_4$, depending only on the mole fractions, are obtained by identification
with the corresponding coefficients which show up in the virial series. Specifically, in the present case one has
\begin{subequations}
\begin{equation}
c_{1} = B^* - 2,\quad c_{2} = C^*-2B^*+1,
\end{equation}
\begin{equation}
c_{3} = D^*-2C^*+B^*,\quad
c_{4} =E^*-2D^*+C^*,
\end{equation}
\end{subequations}
where $D^*\equiv D/\xi^3$ and $E^*\equiv E/\xi^4$.
In turn, the excess Helmholtz free energy per particle  associated with Eq.\ \eqref{rve}  is given by
\beq
\label{FEN-RVE}
\beta a^{\text{RVE}}_{\text{ex}}(\rho)=
\frac{\eta(\alpha_0+\alpha_1\eta+\alpha_2\eta^2)}{1-\eta}-\alpha_L \ln(1-\eta),
\eeq
where $\alpha_0=1+c_1+c_2+2c_3+3c_4$, $\alpha_1=-\left(c_3+\frac{3}{2}c_4\right)$, $\alpha_2=-\frac{1}{2}c_4$, and $\alpha_L=1-c_2-2c_3-3c_4$.
It must be emphasized that, while the H and SHY approximations produce (approximate) expressions of the fourth and fifth virial coefficients  for any composition of the mixture [see Eqs.\ \eqref{virH} and \eqref{virSHY}], the RVE approximation requires the input of the empirical values of those coefficients for each particular mixture.

Regardless of the theory used,  the chemical potential of species $i$ can be obtained from the knowledge of the free energy as a function of the partial densities $\{\rho_i\}$ through the thermodynamic relation
\beq
\mu_i=\left(\frac{\partial\rho a}{\partial \rho_i}\right)_{\rho_{j\neq i}}.
\eeq
The fundamental equation of thermodynamics is equivalent to\cite{S16}
\beq
\beta a+Z=\sum_i x_i \beta \mu_i .
\eeq
In the special case of a binary mixture, it is more convenient to see the free energy $a$ as a function of $\rho$ and $x_1$ [as done in Eqs.\  \eqref{aid}, \eqref{2.8}, \eqref{FEN-SYH}, and \eqref{FEN-RVE}] rather than as a function of $\rho_1$ and $\rho_2$. In that case,
\begin{subequations}
\beq
\beta \mu_1=Z+\beta a+x_2 \left(\frac{\partial\beta a}{\partial x_1}\right)_{\rho},
\eeq
\beq
\beta \mu_2=Z+\beta a-x_1 \left(\frac{\partial\beta a}{\partial x_1}\right)_{\rho}.
\eeq
\end{subequations}

\section{Results}
\label{sec3}
\subsection{Partial virial coefficients}
\label{sec3A}

\begin{figure}
\includegraphics[width=.9\columnwidth]{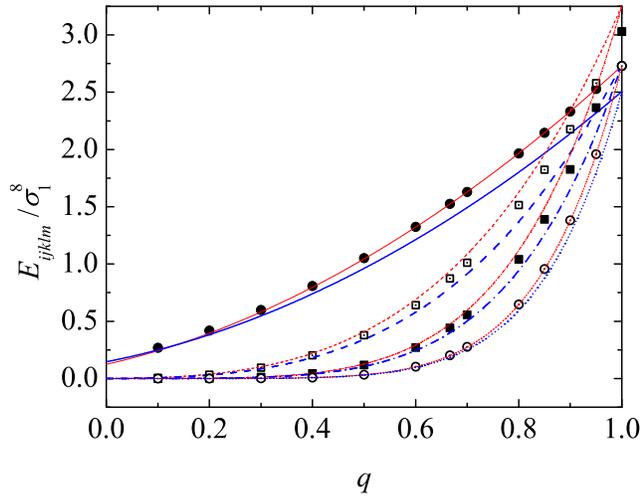}
\caption{Plot of the composition-independent fifth virial coefficients  $E_{11112}/\sigma_1^6$ (solid circles and solid lines), $E_{11122}/\sigma_1^6$ (open squares and dashed lines), $E_{11222}/\sigma_1^6$ (solid squares and dash-dotted lines), and $E_{12222}/\sigma_1^6$ (open circles and dotted lines) versus the size ratio $q=\sigma_2/\sigma_1$ for a nonadditivity parameter $\Delta=0.1$. The symbols are our MC data, while the thin red lines and the thick blue lines  correspond to Hamad's proposal,
Eq.\ \eqref{vir5H}, and  to the SHY proposal, Eq.\ \eqref{5D.1}, respectively.
\label{fig:Delta_01}}
\end{figure}

\begin{figure}
\includegraphics[width=.9\columnwidth]{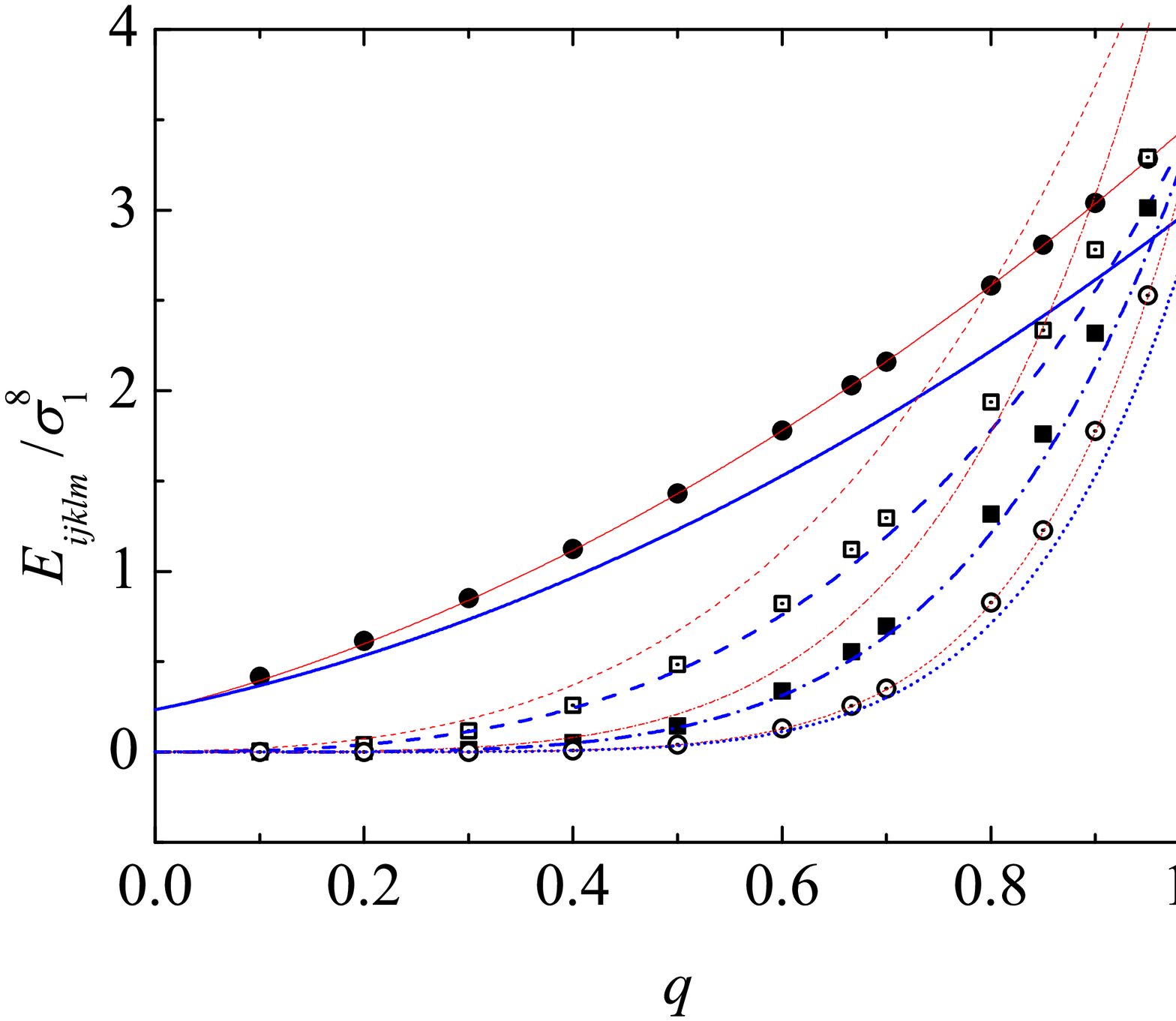}
\caption{Same as in Fig.\ \ref{fig:Delta_01}, but for $\Delta=0.2$.
\label{fig:Delta_02}}
\end{figure}

\begin{figure}
\includegraphics[width=.9\columnwidth]{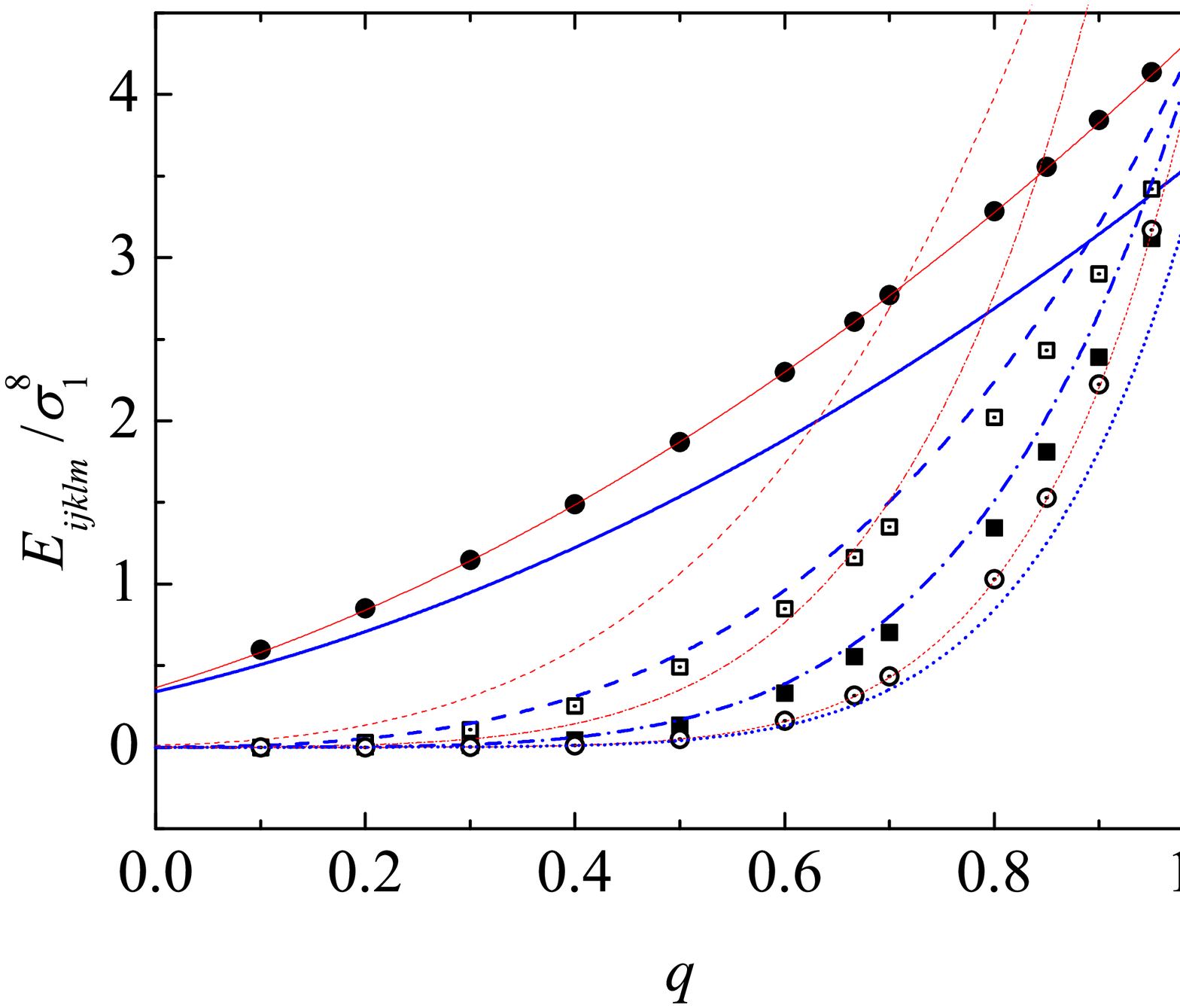}
\caption{Same as in Fig.\ \ref{fig:Delta_01}, but for $\Delta=0.3$.
\label{fig:Delta_03}}
\end{figure}

\begin{figure}
\includegraphics[width=.9\columnwidth]{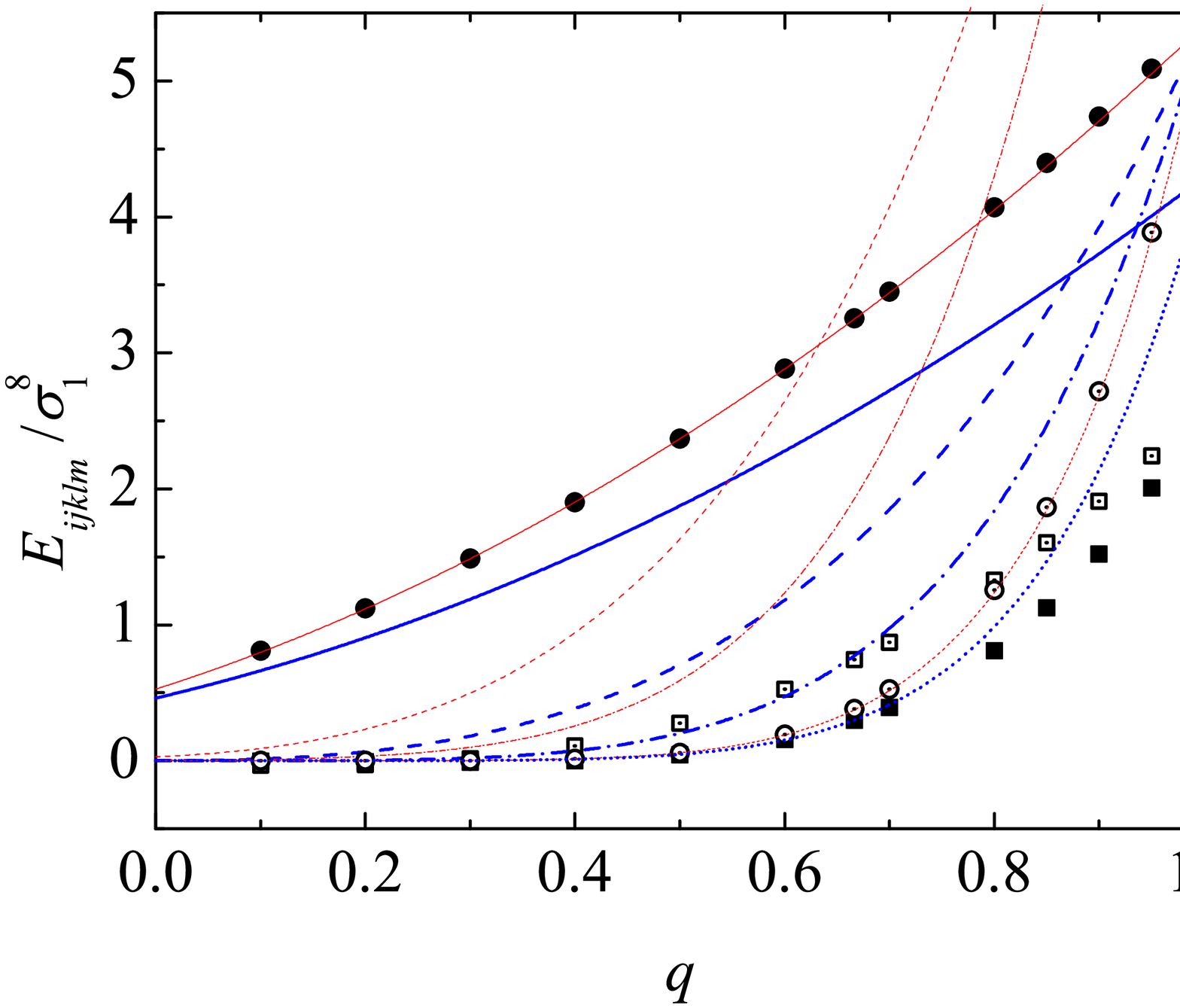}
\caption{Same as in Fig.\ \ref{fig:Delta_01}, but for $\Delta=0.4$.
\label{fig:Delta_04}}
\end{figure}

\begin{figure}
\includegraphics[width=.9\columnwidth]{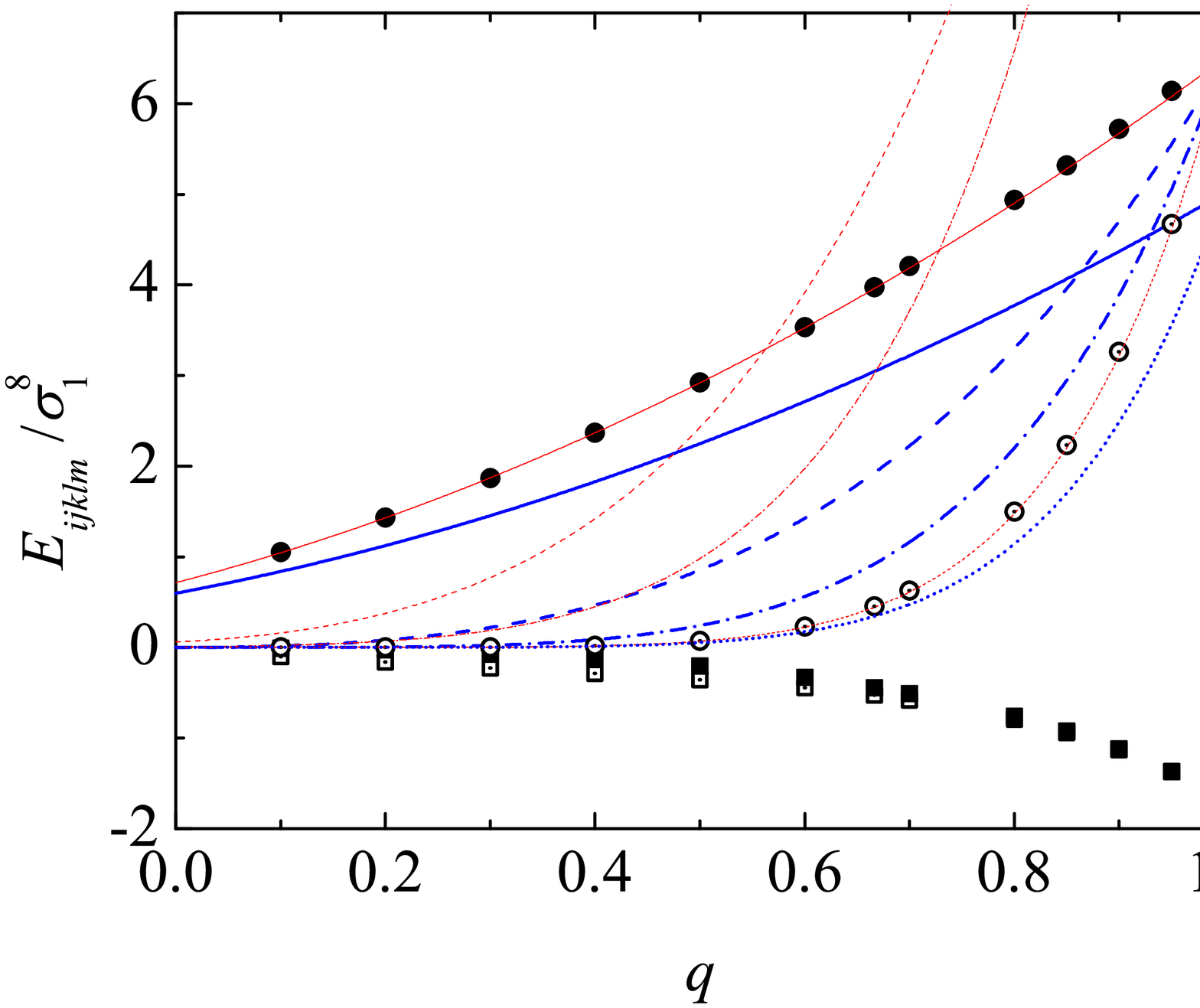}
\caption{Same as in Fig.\ \ref{fig:Delta_01}, but for $\Delta=0.5$.
\label{fig:Delta_05}}
\end{figure}

\begin{figure}
\includegraphics[width=.9\columnwidth]{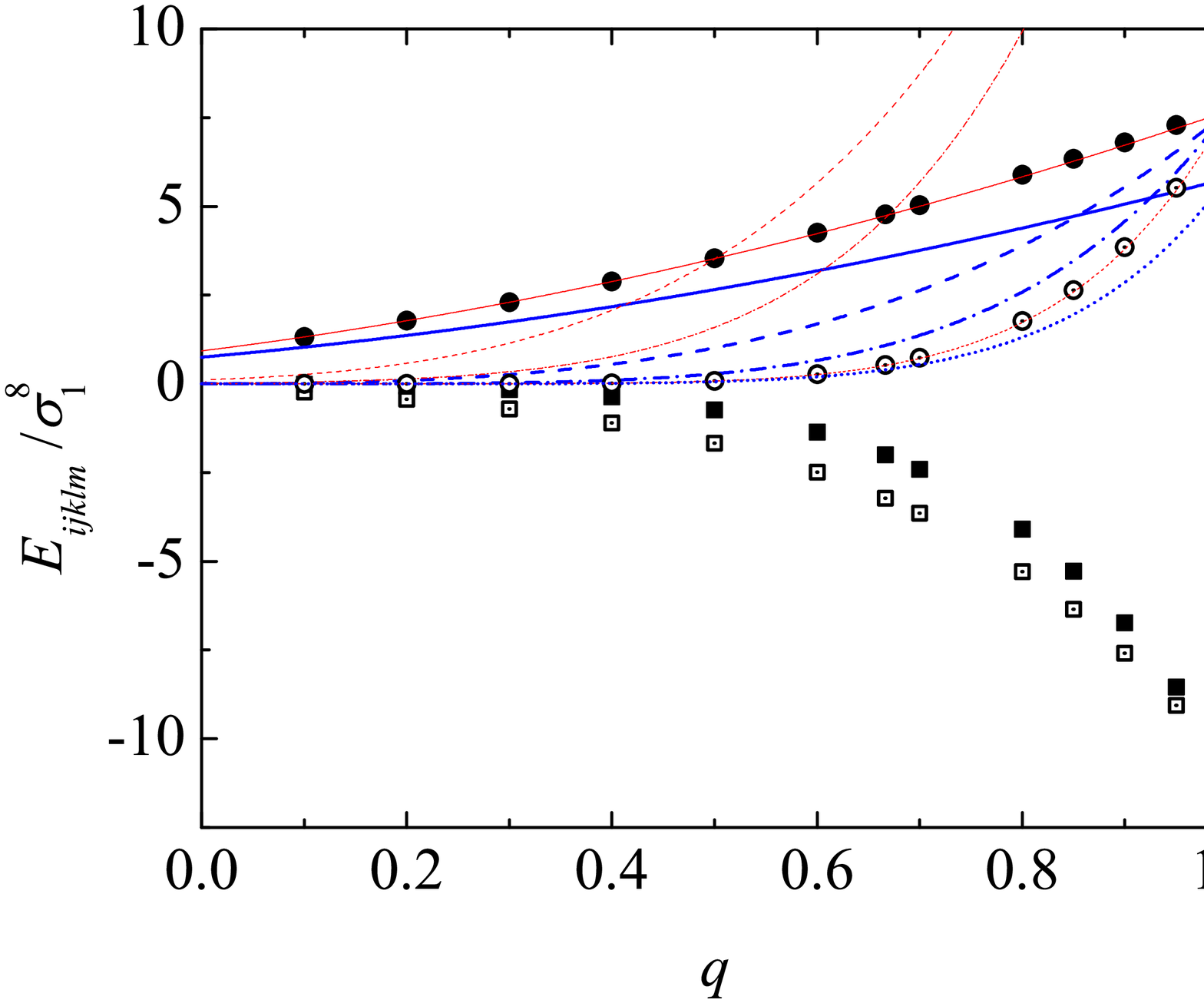}
\caption{Same as in Fig.\ \ref{fig:Delta_01}, but for $\Delta=0.6$.
\label{fig:Delta_06}}
\end{figure}

\begin{figure}
\includegraphics[width=.9\columnwidth]{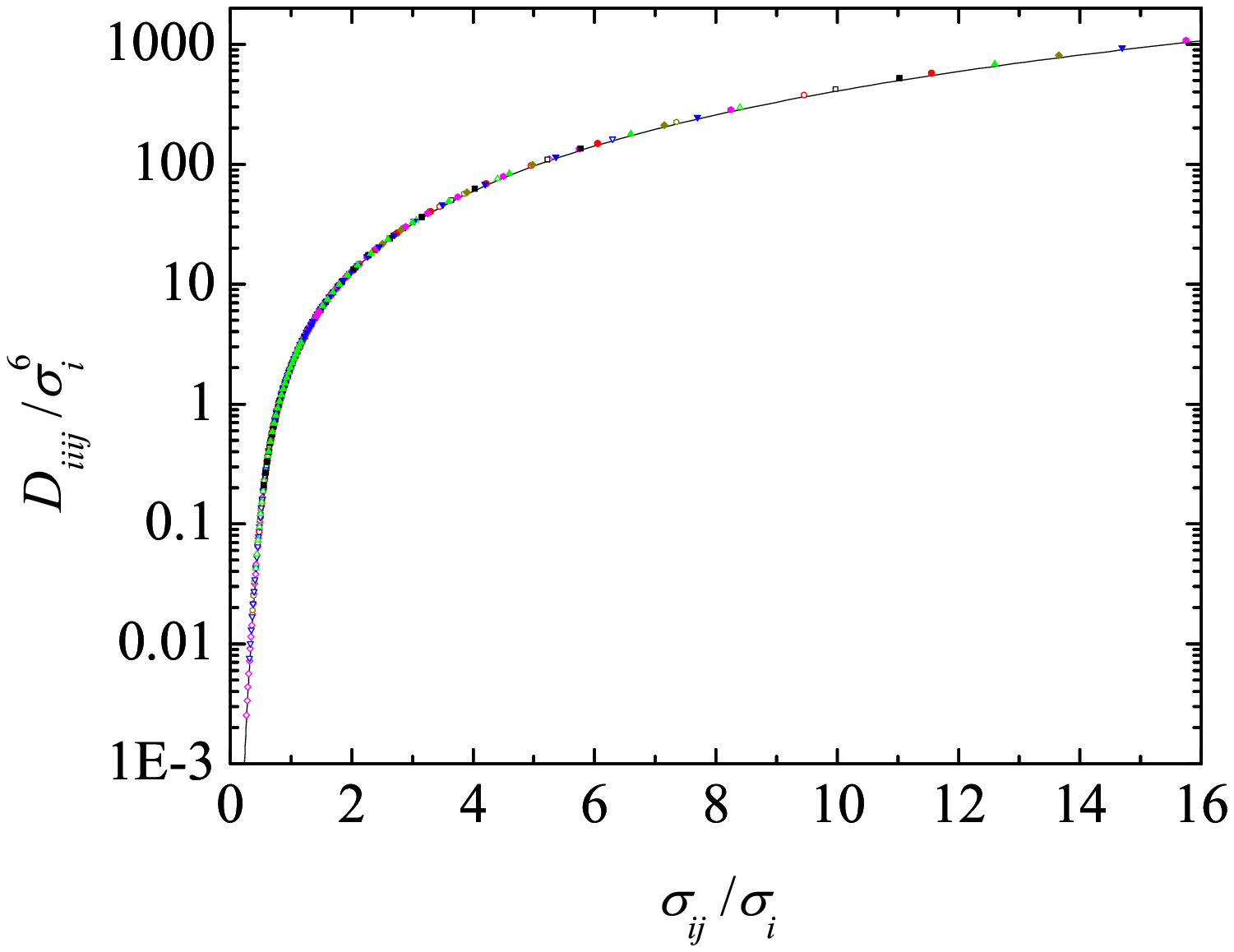}
\caption{Plot of the composition-independent fourth virial coefficient $D_{iiij}/\sigma_i^6$ as a function of the size ratio $\sigma_{ij}/\sigma_i$. The symbols correspond to the values  of  $D_{1112}$ and $D_{1222}$ reported in Ref.\ \ocite{SSYH12} for  $19\times 12=228$ different $(q,\Delta)$ pairs. The line represents  function \eqref{D_Hamad}.
\label{fig:Collapse_D}}
\end{figure}

\begin{figure}
\includegraphics[width=.9\columnwidth]{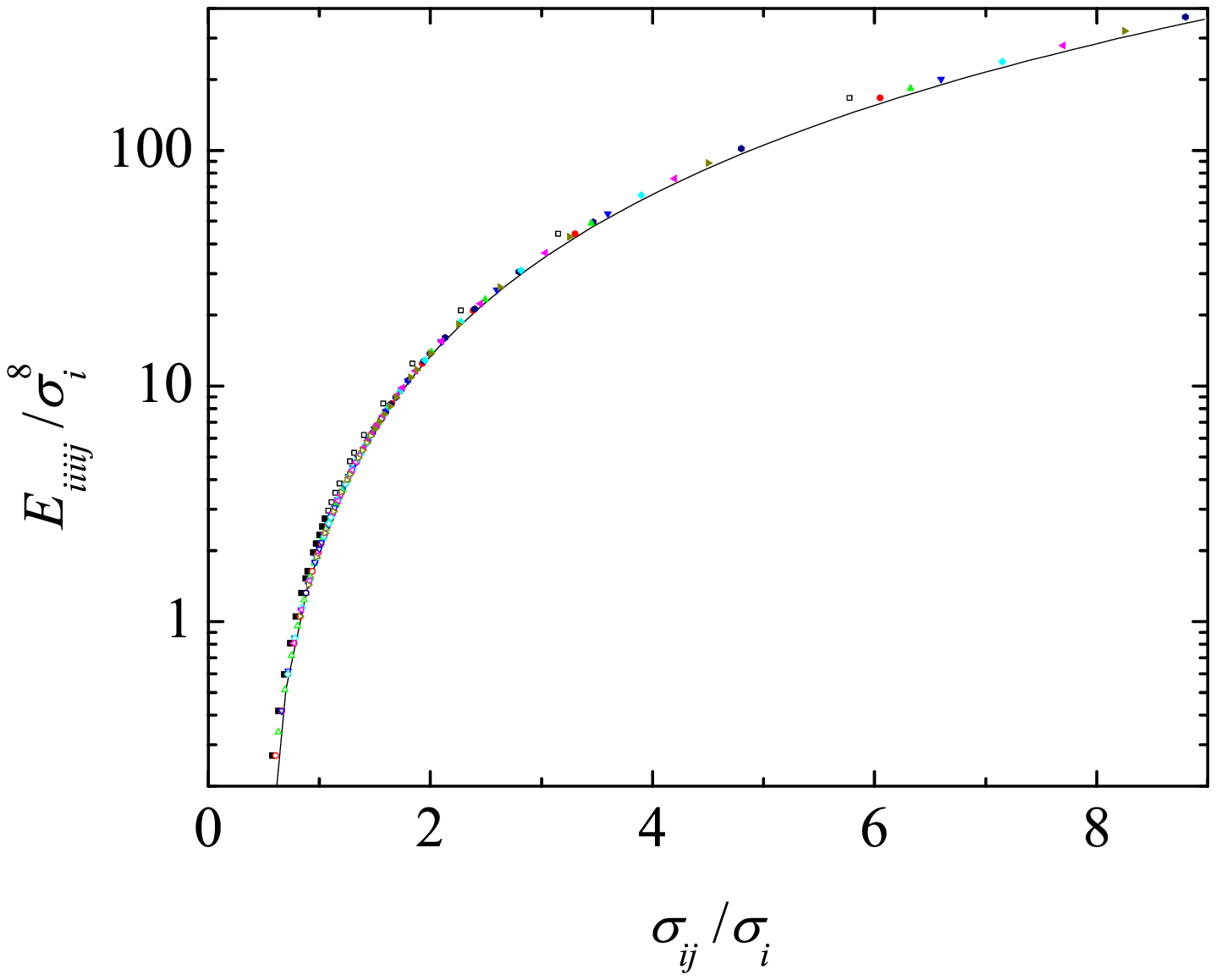}
\caption{Plot of the composition-independent fifth virial coefficient $E_{iiiij}/\sigma_i^8$ as a function of the size ratio $\sigma_{ij}/\sigma_i$. The symbols correspond to the values  of  $E_{11112}$ and $E_{12222}$ reported in this work for  $13\times 8=104$ different $(q,\Delta)$ pairs. The line represents function \eqref{E_Hamad}.
\label{fig:Collapse_E}}
\end{figure}

\begin{figure}
\includegraphics[width=\columnwidth]{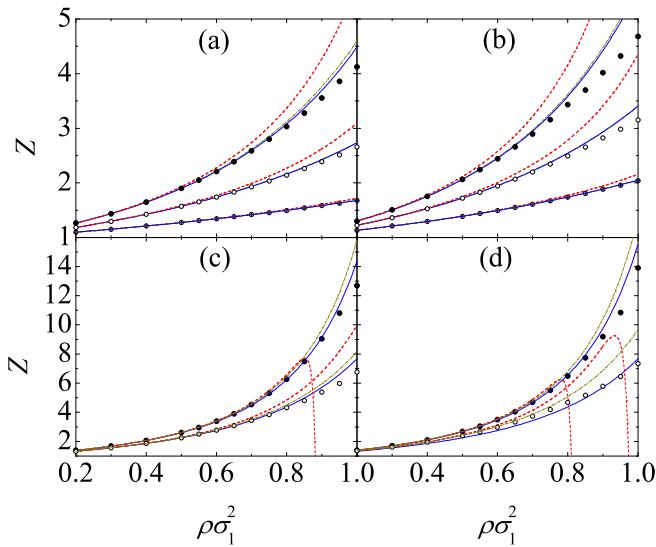}
\caption{Plot of the compressibility factor $Z$ as a function of density for size ratios $q=0.4$ [panels (a) and (c)] and $q=0.5$ [panels (b) and (d)], in all the cases with a nonadditivity parameter $\Delta=0.3$. The mole fractions are,   from  bottom to top, $x_1=0.1$, $0.3$, $0.5$ [panels (a) and (b)] and $x_1=0.7$, $0.9$ [panels (c) and (d)]. The symbols correspond to our MC simulation data, while the dashed, solid, and dash-dotted lines correspond to the H [Eq.\ \protect\eqref{2.7}], SHY [Eq.\ \protect\eqref{new2}], and RVE [Eq.\ \protect\eqref{rve}] approximations, respectively. Note that the SHY and RVE curves for $x_1=0.1$ and $0.3$ are hardly distinguishable. Uncertainties in our MC results
are smaller than the symbol size.
\label{fig:Z_q04_05}}
\end{figure}

\begin{figure}
\includegraphics[width=\columnwidth]{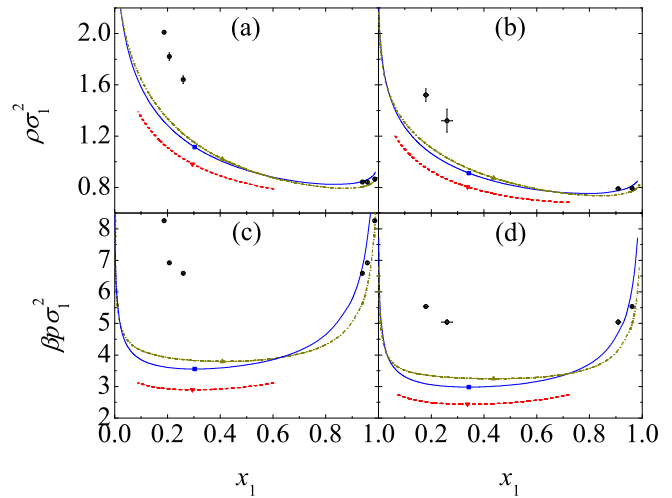}
\caption{Fluid-fluid coexistence curves in the  reduced density vs composition plane [panels (a) and (b)] and reduced pressure vs composition plane [panels (c) and (d)] for size ratios $q=0.4$ [panels (a) and (c)] and $q=0.5$ [panels (b) and (d)], in both cases with a nonadditivity parameter $\Delta=0.3$.
The circles (with horizontal and vertical error bars) correspond to our GEMC simulation data, while the dashed, solid, and dash-dotted lines correspond to the H [Eq.\ \protect\eqref{2.7}], SHY [Eq.\ \protect\eqref{new2}], and RVE [Eq.\ \protect\eqref{rve}] approximations, respectively. The filled symbols on the theoretical curves indicate the location of the corresponding critical points. The points to the right (left) of the critical point represent phase I (II), i.e., the phase rich (poor) in big disks. \label{fig1demixing}}
\end{figure}

\begin{figure}
\includegraphics[width=\columnwidth]{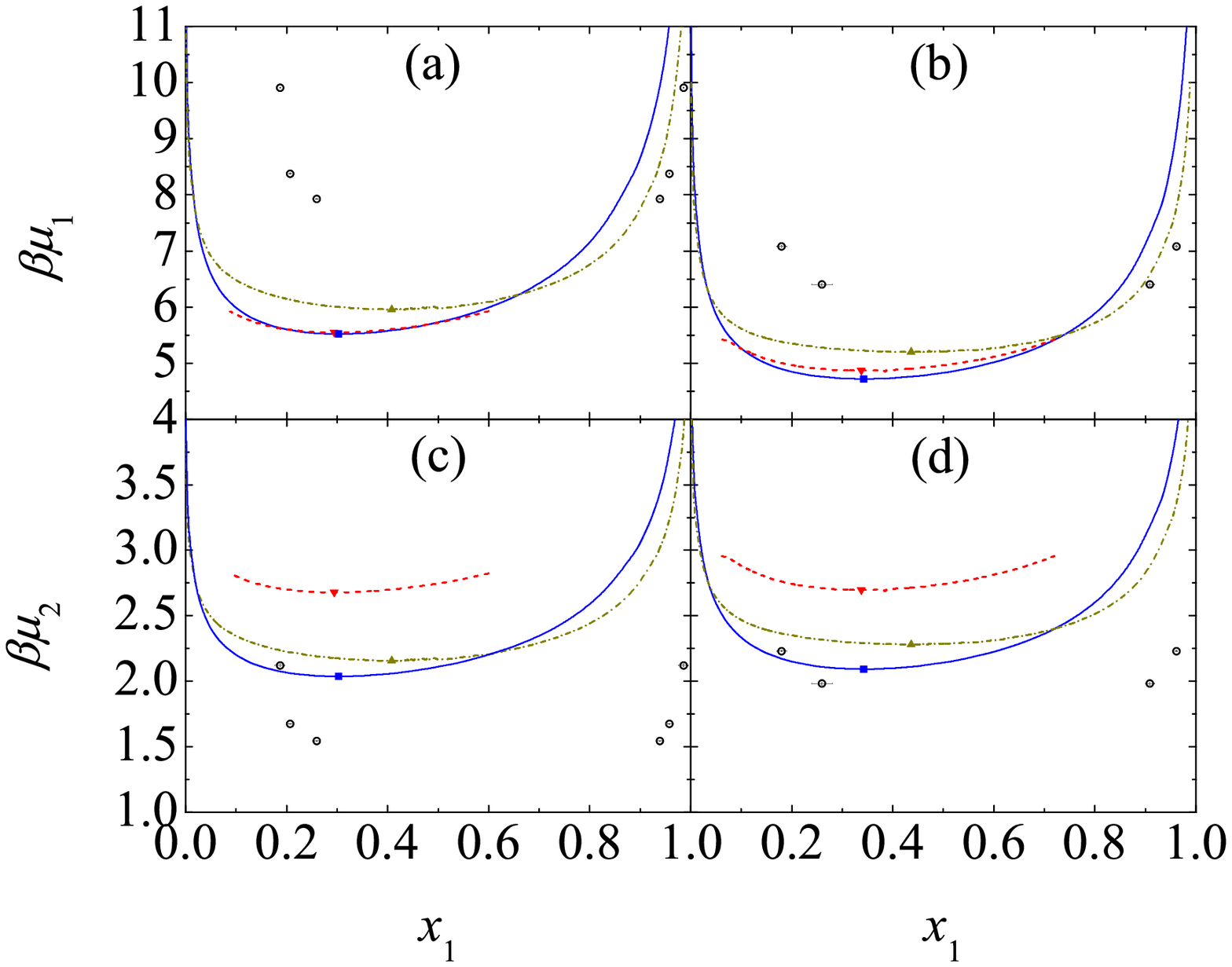}
\caption{Same as in Fig.\ \ref{fig1demixing} but for  the  reduced chemical potential of each species vs composition plane. \label{fig2demixing}}
\end{figure}

In a previous paper, some of us reported\cite{SSYH12} MC calculations of the composition-independent partial fourth virial coefficients of asymmetric NAHD mixtures $D_{1112}$, $D_{1122}$, and $D_{1222}$ over a rather wide range of size ratios $q=\sigma_2/\sigma_1$ and values of the nonadditivity parameter $\Delta$. Those data were complementary to the ones reported earlier\cite{S11} by one of us for symmetric mixtures. Here we continue with our efforts to provide additional information on the virial coefficients of NAHD mixtures and report the numerical values of the partial fifth virial coefficients for a range of size ratios and  positive nonadditivities.

As in the case of the fourth virial coefficients, in order to evaluate the irreducible cluster integrals which enter the expression of the composition-independent coefficients $E_{ijk\ell m}$ (see Table \ref{table_Eijklm}), we used a standard MC integration procedure \cite{SFG96}, although more efficient techniques to compute higher order virial coefficients have been proposed.\cite{LKM05,W13,ZP14}
Our algorithm produces a significant set of configurations which are compatible with the Mayer graph one wants to evaluate. We first fix particle $1$ of species $i=1$ at the origin and sequentially deposit the remaining four particles (of species $i=1$ or $i=2$, depending on the diagram in Table \ref{table_Eijklm}) at random positions but in such a way that particle $\alpha=2,3,4,5$ overlaps with particle $\alpha-1$. This procedure generates an open chain of overlapping particles which is taken as a ``trial configuration''. In a ``successful configuration''  the residual cross-linked ``bonds''  present in the Mayer graph that is being calculated are retrieved. The ratio of the number of successful configurations ($N_s$) to the total number of trial configurations ($N_t$) yields asymptotically the value of the cluster integral relative to that of the open-chain graph which, in turn, is trivially related to a product of the partial second-order virial coefficients $B_{ij}$.\cite{BN86,B92} The numerical accuracy of the MC results obviously depends on the total number of trial configurations. The relative error on the evaluation of a given coefficient is  estimated as\cite{K77}
\begin{equation}
\text{error} = \left[ \frac {J(J-1)}{N_t} \right]^{1/2},
\label{s:error}
\end{equation}
where  $J$ is the number of cluster integrals contributing to that coefficient.
However, as a result of the accumulation of statistically independent errors, the global uncertainty affecting the partial virial coefficients is higher than the error estimated for each cluster integral that enters the expression of $E_{ijk\ell m}$.
A typical MC run consisted of $4\times10^9$--$5\times 10^{11}$ independent moves, depending on the value of $q$.
In the latter case, a typical run lasted approximately $25$ hours on a quad-processor computer based on Intel Xeon $3.4$ GHz processors equipped with $32$ GB RAM. Independent pseudorandom numbers were produced by adopting the Mersenne Twister MT19937 pseudorandom number generator, which has a very long period ($2^{19\,937}-1$).
The error on each cluster integral, as estimated through Eq.\ \eqref{s:error}, turned out to be systematically less than $0.05\%$, with a cumulative uncertainty on the partial virial coefficients lower than $0.5\%$.
By repeating  the evaluation of some partial virial coefficients with independent MC configurations, summing up the cluster integrands for each
configuration, and using the standard deviation of the block averages, we have checked that the uncertainty again remains lower than $0.5\%$, in agreement with the error estimate from  Eq.\ \eqref{s:error}.

The numerical values of $E_{11112}$, $E_{11122}$, $E_{11222}$, and $E_{12222}$ (in units of $\sigma_1^6$) for $\Delta=0.05$, $0.1$, $0.15$, $0.2$, $0.3$, $0.4$, $0.5$,  $0.6$, and  $13$ values of $q$ within the range $0.1\leq q \leq 1$ are presented in Tables I--VIII in the supplementary material to this paper. The values corresponding to $q=1$ (symmetric mixtures) included in the supplemental material  were already reported in Ref.\ \ocite{FPPS14}. Note also that the case of additive hard-disk mixtures was considered by Wheatley in Ref.\ \ocite{W98c}.

In Figs.\ \ref{fig:Delta_01}--\ref{fig:Delta_06} we plot our MC values of the partial fifth virial coefficients as functions of $q$ for $\Delta=0.1$--$0.6$, and assess the performance of the H and SHY approximations
against the MC data.
We can observe that the H prescription provides excellent estimates of $E_{11112}$ and $E_{12222}$ for the whole range of $q$ and for all the values of the nonadditivity parameter considered. The SHY approximation gives reasonable results for those two coefficients, but it is much less accurate than the H one, especially as one approaches the limit of symmetric mixtures ($q\to 1$).

 With respect to the coefficients $E_{11122}$ and $E_{11222}$, both theories provide rather good predictions for relatively small nonadditivity, namely, $\Delta=0.05$ (not shown), $0.1$, and $0.15$ (not shown). For moderate nonadditivity (say $\Delta=0.2$ and $0.3$), the SHY approximation provides fair results, while the H one becomes very poor. For large nonadditivity ($\Delta\geq 0.4$) neither of the two theories compare well with the MC values of $E_{11122}$ and $E_{11222}$, which  clearly take negative values if $\Delta \geq 0.5$. Similar results were observed in Ref.\ \ocite{SSYH12} in the case of the coefficients $D_{ijk\ell}$ for positive nonadditivities.

It is interesting to note that, by accident, in the symmetric limit $q\to 1$ the SHY values of $E_{11122}=E_{11222}$ are practically indistinguishable from the H values of $E_{11112}=E_{12222}$ for all $\Delta>0$.

Before closing this subsection, it is worthwhile testing the internal consistency of our numerical evaluation of the partial coefficients $D_{1112}$, $D_{1222}$, $E_{11112}$, and $E_{12222}$. On physical grounds, those coefficients must depend only on $\sigma_{12}$ and either $\sigma_1$ (in the case of $D_{1112}$ and $E_{11112}$) or $\sigma_2$ (in the case of $D_{1222}$ and $E_{12222}$), regardless of whether the mixture is additive or not. More specifically,
\beq
\label{D1112&E11112}
D_{iiij}=\sigma_i^6 \mathcal{D}\left(\frac{\sigma_{ij}}{\sigma_i}\right),\quad E_{iiiij}=\sigma_i^8 \mathcal{E}\left(\frac{\sigma_{ij}}{\sigma_i}\right).
\eeq
Therefore, a plot of $D_{1112}/\sigma_1^6$ and $D_{1222}/\sigma_2^6$ versus  $\sigma_{12}/\sigma_1$ and $\sigma_{12}/\sigma_2$, respectively, for different pairs of $q$ and $\Delta$, must collapse into a single curve. The same must occur for $E_{11112}/\sigma_1^8$ and $E_{12222}/\sigma_2^8$.
This is shown in Figs.\ \ref{fig:Collapse_D} and \ref{fig:Collapse_E}, where an excellent degree of internal consistency is found. Those figures also include the theoretical predictions stemming from Eqs.\ \eqref{virH}, namely,
\begin{subequations}
\beq
\mathcal{D}^{\text{H}}(x)=\left(\frac{\pi}{4}\right)^3\frac{b_4}{2}\left\{\frac{8\mathcal{G}(x)}{\pi b_3}+x^2\left[\frac{8\mathcal{H}(x)}{\pi b_3}\right]^2\right\},
\label{D_Hamad}
\eeq
\beq
\mathcal{E}^{\text{H}}(x)=\left(\frac{\pi}{4}\right)^4\frac{b_5}{5}\left\{{3}\frac{8\mathcal{G}(x)}{\pi b_3}+{2}x^2\left[\frac{8\mathcal{H}(x)}{\pi b_3}\right]^3\right\}.
\label{E_Hamad}
\eeq
\end{subequations}
As already observed in Ref.\ \ocite{SSYH12} and in Figs.\ \ref{fig:Delta_01}--\ref{fig:Delta_06}, an excellent performance of the H predictions for $D_{iiij}$ and $E_{iiiij}$ is confirmed by Figs.\  \ref{fig:Collapse_D} and \ref{fig:Collapse_E}. In contrast, we have found that the SHY expressions obtained from Eqs.\ \eqref{virSHY}  do not conform to the scaling relations \eqref{D1112&E11112}.

\subsection{Compressibility factor}
\label{sec3B}

We have performed canonical MC simulations to measure the compressibility factor as a function of density for NAHD mixtures with $\Delta=0.3$, two size ratios ($q=0.4$ and $0.5$), and five mole fractions ($x_1=0.1$, $0.3$, $0.5$, $0.7$, and $0.9$). The results are contained in Tables IX and X of the supplementary material and displayed graphically in Fig.\ \ref{fig:Z_q04_05}. In this figure we have also included the corresponding values obtained from the three theoretical compressibility factors $Z^{\text{H}}(\rho)$, $Z^{\text{SHY}}(\rho)$, and $Z^{\text{RVE}}(\rho)$  as given by Eqs.\ \eqref{2.7}, \eqref{new2}, and \eqref{rve}, respectively. As explained before, the RVE compressibility factor needs to be complemented with the empirical virial coefficients. As for the H and SHY EoSs, we have taken for $Z^{\text{pure}}(\eta)$ the accurate compressibility factor corresponding to the EoS proposed by Luding,\cite{L01b,LS01,LS04} namely,
\beq
   Z^{\text{pure}}(\eta)=\frac{1+\eta^2/8}{(1-\eta)^2}-\frac{\eta^4}{64(1-\eta)^4}.
    \label{Luding}
    \eeq

One can clearly see from Fig.\ \ref{fig:Z_q04_05} that in this case the best performance undoubtedly corresponds to $Z^{\text{SHY}}$, which only deviates from the MC data for the highest densities. It is also remarkable that the SHY compressibility factor is as accurate and even in some instances more accurate than the RVE, since the latter involves the first five exact virial coefficient while the former only yields exactly the first three virial coefficients.
This shows that the RVE EoS \eqref{rve} is more sensitive than the SHY EoS \eqref{new2} to the positive nonadditivity consequence of the effective occupied
volume fraction being larger than the nominal packing fraction $\eta$. In fact, as we observed in Ref.\ \ocite{SSYH12}, the RVE EoS performs better than the SHY EoS in the case of negative nonadditivity.

The poorest agreement with the MC data corresponds to the Hamad compressibility factor, which may even become negative after a certain density. This feature is the result of  the combination of the special density dependence in $Z^{\text{H}}(\rho)$ and the breakdown of the Luding EoS for extremely high packing fractions. Note that Hamad's EoS makes use of $Z^{\text{pure}}$ at three effective packing fractions, namely, $\eta X_{11}$, $\eta X_{12}$, and $\eta X_{22}$ [see Eq.\ \eqref{2.7}]. As an illustration, let us consider the mixture with $\Delta=0.3$, $q=0.5$, and $x_1=0.9$ [top curve in Fig.\ \ref{fig:Z_q04_05}(d)].  For the chosen mixture, in order to get $Z^{\text{H}}(\rho)$ at a given $\rho$, one has to know $Z^{\text{pure}}$ at $\eta=\rho \xi X_{22}=1.18334\rho $. Since the maximum value of the packing fraction in the one-component fluid is $\eta=\sqrt{3}\pi/6=0.9069$, the corresponding maximum value of the density that one might consider in the mixture would be $\rho=0.9069/1.18334=0.7664$. On the other hand, Luding's EoS presents a local maximum at $\eta=0.8649$ and becomes negative after $\eta=0.9029$. Hence, $Z^{\text{H}}(\rho)$ may not be obtained beyond $\rho=0.8649/1.18334=0.7309$.

Once we have assessed the merits of the theoretical proposals for the EoS of NAHD mixtures with respect to their predictions of the virial coefficients and from the comparison with the MC values for the compressibility factors, we now turn to the problem of demixing in such mixtures.

\subsection{Demixing}

We have performed GEMC simulations in order to determine the fluid-fluid coexistence curve of a few binary asymmetric NAHD mixtures. The detailed description of the method will be omitted since it is presently a well-established procedure and a thorough account of it can be found for instance in Refs.\ \onlinecite{P92,FS02}. Nevertheless, we should point out that in order to simulate two coexisting phases one performs a MC simulation with two coupled boxes, each one containing a homogeneous phase. Then, three different types of moves are performed to allow for thermal, mechanical, and chemical equilibration. The first one is the usual Metropolis displacement of particles in each box separately. The second type of move is the change of volume of the two regions. The third type is the exchange of particles between the two boxes. One should add here that, due to the high asymmetry and the high nonadditivity, in the region of phase separation the particle exchange rate becomes rather slow and so the equilibration of the system requires very long runs.
Simulation boxes were filled with an initial number of particles ranging between $1500$--$2000$ particles per box, depending on the desired initial total density.
The chemical potentials of the two species were calculated by using  Widom's  particle insertion method,\cite{W63b} and the appropriate formula for the Gibbs ensemble.\cite{FS02} The attempted particle sweeps between the two boxes of each GEMC cycle were used to sample the energy landscape of the system containing the inserted particle.

We have considered the same two NAHD mixtures as in Sec.\ \ref{sec3B}, namely, those with $\Delta=0.3$ and size ratios $q=0.4$ and $q=0.5$. For these very asymmetric NAHD mixtures the GEMC method becomes extremely difficult and, as a consequence, only a few points of the coexistence curve could be accurately determined.  On the other hand, those points provide a reasonable idea about the location of the phase coexistence curve.
Table XI of the supplementary material contains the GEMC simulation values of the coordinates (composition, $x_1$, reduced density, $\rho \sigma_1^2$, reduced pressure, $\beta p \sigma_1^2$, and reduced chemical potentials, $\beta\mu_i$) of three pairs of points ($q=0.4$) and two pairs of points ($q=0.5$) in the binodals.

On the theoretical side, we have computed the same binodals curves, as predicted by the  H, SHY, and RVE approaches (in the latter case by injecting again the empirical virial coefficients), by imposing the equality of the pressure $p$ and the chemical potentials of both species $\mu_1$ and $\mu_2$, respectively, in the two fluid phases.

In Figs.\ \ref{fig1demixing}  and \ref{fig2demixing}, we display the results of our GEMC computations   and  include also the theoretical results for comparison.  It is clear that all the theories heavily underestimate the values of the density, the pressure, and the chemical potential of the big disks at coexistence in the fluid  phase which is rich in the small disks, while they overestimate the chemical potential of the small disks in the phase rich in the big disks.  However, both the SHY and RVE theories are rather accurate in predicting the density and pressure of the phase rich in the big disks. We note that in the case of the H EoS, due to the same limitations mentioned above in connection with Fig.\ \ref{fig:Z_q04_05}, numerical instabilities arose that prevented us from covering the whole composition range. In any event, this H EoS exhibits the worst performance.

It may seem paradoxical that, while  both the SHY and RVE approaches produce good outputs for low $x_1$ values at a
single phase, as shown in Fig.\ \ref{fig:Z_q04_05},  they produce large underestimates of the number density for the
low $x_1$ branch of the coexistence curve in Fig.\ \ref{fig1demixing}. The explanation is two-fold. On the one hand, the coexistence conditions of equal pressure and chemical potentials in two very disparate phases is a much more delicate and stringent test of a theory than the equation of state in a mixed phase. On the other hand, the coexisting phase  rich in small disks has in general a very large number density, even twice as large as the maximum value considered in  Fig.\ \ref{fig:Z_q04_05}.

\section{Concluding remarks}
\label{sec4}

In this paper we have reported MC calculations of the fifth virial coefficients of asymmetric NAHD mixtures for a range of size ratios $q$ and  various values of the nonadditivity parameter $\Delta$. These results complement those reported earlier\cite{S11,SSYH12} and were used to assess the predictions of the same coefficients coming from the H\cite{AEH99} and the SHY.\cite{SHY05} proposals. They were also used to derive the RVE for NAHD mixtures truncated at the fifth order. While the H EoS leads to extremely accurate predictions for  the partial fourth and fifth virial coefficients $D_{iiij}$ and $E_{iiiij}$, respectively, it may grossly err in the case of others. In this sense, although the SHY proposal may also lead in its predictions to percent deviations of the order of 20\%, it appears to be a reasonably simple overall compromise.

The three theoretical EoSs for the mixtures were further tested against MC simulation data of the compressibility factor. In this case, it is remarkable that the SHY proposal yields the best overall performance, with accuracy as good or even better than the RVE EoS, given the fact that it is relatively simple and does not rely upon the empirical knowledge of virial coefficients beyond the third. On the other hand, the limitations of the H EoS make it the worst approximation in this instance.

Concerning the demixing problem, while all the theories are unable to produce accurate values of the density and the pressure of the coexisting phase rich in small disks, the other coexisting phase is well accounted for by the SHY and RVE proposals.  Thus, again we find that the SHY EoS appears at this stage to be a good choice in view of its relative simplicity and its independence of the external numerical determination of virial coefficients.  In any case, the point we want to stress finally is that more theoretical efforts to cope correctly with demixing in NAHD mixtures are called for.

\section*{Supplementary Material}
See supplementary material for  tables containing the results of our computations of the composition-independent fifth virial coefficients, compressibility factor, and binodal curves  of asymmetric nonadditive hard-disk mixtures.

\begin{acknowledgments}
One of us (G.F.) wishes to thank the Department of Mathematics and Computer Science, Physics and Earth Sciences at
the University of Messina for kindly providing computing
facilities.
G.P. thanks the National Research Foundation of South Africa for financial support with Grant No.\ 96463 (Blue Skies) and acknowledges Mr. A. Roth for some preliminary GEMC calculations.
Two of us (A.S. and S.B.Y) acknowledge the financial support of the
Ministerio de Econom\'ia y Competitividad (Spain) through Grant No.\ FIS2016-76359-P and  the Junta de Extremadura (Spain) through Grant No.\ GR15104, both partially financed by ``Fondo Europeo de Desarrollo Regional'' funds. M.L.H. thanks CONACYT for a sabbatical grant.
\end{acknowledgments}


\begin{thebibliography}{53}
\expandafter\ifx\csname natexlab\endcsname\relax\def\natexlab#1{#1}\fi
\expandafter\ifx\csname bibnamefont\endcsname\relax
  \def\bibnamefont#1{#1}\fi
\expandafter\ifx\csname bibfnamefont\endcsname\relax
  \def\bibfnamefont#1{#1}\fi
\expandafter\ifx\csname citenamefont\endcsname\relax
  \def\citenamefont#1{#1}\fi
\expandafter\ifx\csname url\endcsname\relax
  \def\url#1{\texttt{#1}}\fi
\expandafter\ifx\csname urlprefix\endcsname\relax\def\urlprefix{URL }\fi
\providecommand{\bibinfo}[2]{#2}
\providecommand{\eprint}[2][]{\url{#2}}

\bibitem[{\citenamefont{Dickinson}(1977)}]{D77}
\bibinfo{author}{\bibfnamefont{E.}~\bibnamefont{Dickinson}},
  \bibinfo{journal}{Mol. Phys.} \textbf{\bibinfo{volume}{33}},
  \bibinfo{pages}{1463} (\bibinfo{year}{1977}).

\bibitem[{\citenamefont{Dickinson}(1979)}]{D79}
\bibinfo{author}{\bibfnamefont{E.}~\bibnamefont{Dickinson}},
  \bibinfo{journal}{Chem. Phys. Lett.} \textbf{\bibinfo{volume}{66}},
  \bibinfo{pages}{500} (\bibinfo{year}{1979}).

\bibitem[{\citenamefont{Dickinson}(1980)}]{D80}
\bibinfo{author}{\bibfnamefont{E.}~\bibnamefont{Dickinson}},
  \bibinfo{journal}{J. Chem. Soc. Faraday Trans. 2}
  \textbf{\bibinfo{volume}{76}}, \bibinfo{pages}{1458} (\bibinfo{year}{1980}).

\bibitem[{\citenamefont{Tenne and Bergmann}(1978)}]{TB78}
\bibinfo{author}{\bibfnamefont{R.}~\bibnamefont{Tenne}} \bibnamefont{and}
  \bibinfo{author}{\bibfnamefont{E.}~\bibnamefont{Bergmann}},
  \bibinfo{journal}{Phys. Rev. A} \textbf{\bibinfo{volume}{17}},
  \bibinfo{pages}{2036} (\bibinfo{year}{1978}).

\bibitem[{\citenamefont{Singh and Sinha}(1983)}]{SS83}
\bibinfo{author}{\bibfnamefont{U.~N.} \bibnamefont{Singh}} \bibnamefont{and}
  \bibinfo{author}{\bibfnamefont{S.~K.} \bibnamefont{Sinha}},
  \bibinfo{journal}{Pramana} \textbf{\bibinfo{volume}{20}},
  \bibinfo{pages}{327} (\bibinfo{year}{1983}).

\bibitem[{\citenamefont{Mishra and Sinha}(1985)}]{MS85}
\bibinfo{author}{\bibfnamefont{B.~M.} \bibnamefont{Mishra}} \bibnamefont{and}
  \bibinfo{author}{\bibfnamefont{S.~K.} \bibnamefont{Sinha}},
  \bibinfo{journal}{J. Math. Phys.} \textbf{\bibinfo{volume}{26}},
  \bibinfo{pages}{495} (\bibinfo{year}{1985}).

\bibitem[{\citenamefont{Bearman and Mazo}(1988)}]{BM88}
\bibinfo{author}{\bibfnamefont{R.~J.} \bibnamefont{Bearman}} \bibnamefont{and}
  \bibinfo{author}{\bibfnamefont{R.~M.} \bibnamefont{Mazo}},
  \bibinfo{journal}{J. Chem. Phys.} \textbf{\bibinfo{volume}{88}},
  \bibinfo{pages}{1235} (\bibinfo{year}{1988}).

\bibitem[{\citenamefont{Bearman and Mazo}(1989)}]{BM89}
\bibinfo{author}{\bibfnamefont{R.~J.} \bibnamefont{Bearman}} \bibnamefont{and}
  \bibinfo{author}{\bibfnamefont{R.~M.} \bibnamefont{Mazo}},
  \bibinfo{journal}{J. Chem. Phys.} \textbf{\bibinfo{volume}{91}},
  \bibinfo{pages}{1227} (\bibinfo{year}{1989}).

\bibitem[{\citenamefont{Bearman and Mazo}(1990)}]{BM90}
\bibinfo{author}{\bibfnamefont{R.~J.} \bibnamefont{Bearman}} \bibnamefont{and}
  \bibinfo{author}{\bibfnamefont{R.~M.} \bibnamefont{Mazo}},
  \bibinfo{journal}{J. Chem. Phys.} \textbf{\bibinfo{volume}{93}},
  \bibinfo{pages}{6694} (\bibinfo{year}{1990}).

\bibitem[{\citenamefont{Ehrenberg et~al.}(1990)\citenamefont{Ehrenberg,
  Schaink, and Hoheisel}}]{ESH90}
\bibinfo{author}{\bibfnamefont{V.}~\bibnamefont{Ehrenberg}},
  \bibinfo{author}{\bibfnamefont{H.~M.} \bibnamefont{Schaink}},
  \bibnamefont{and} \bibinfo{author}{\bibfnamefont{C.}~\bibnamefont{Hoheisel}},
  \bibinfo{journal}{Physica A} \textbf{\bibinfo{volume}{169}},
  \bibinfo{pages}{365} (\bibinfo{year}{1990}).

\bibitem[{\citenamefont{Fraser et~al.}(1991)\citenamefont{Fraser, Zuckermann,
  and Mouritsen}}]{FZM91}
\bibinfo{author}{\bibfnamefont{D.~P.} \bibnamefont{Fraser}},
  \bibinfo{author}{\bibfnamefont{M.~J.} \bibnamefont{Zuckermann}},
  \bibnamefont{and} \bibinfo{author}{\bibfnamefont{O.~G.}
  \bibnamefont{Mouritsen}}, \bibinfo{journal}{Phys. Rev. A}
  \textbf{\bibinfo{volume}{43}}, \bibinfo{pages}{6642} (\bibinfo{year}{1991}).

\bibitem[{\citenamefont{Marti and Croset}(1994)}]{MC94}
\bibinfo{author}{\bibfnamefont{C.~C.} \bibnamefont{Marti}} \bibnamefont{and}
  \bibinfo{author}{\bibfnamefont{B.~J.} \bibnamefont{Croset}},
  \bibinfo{journal}{Surf. Sci.} \textbf{\bibinfo{volume}{318}},
  \bibinfo{pages}{229} (\bibinfo{year}{1994}).

\bibitem[{\citenamefont{Nielaba}(1996)}]{N96}
\bibinfo{author}{\bibfnamefont{P.}~\bibnamefont{Nielaba}},
  \bibinfo{journal}{Int. J. Thermophys.} \textbf{\bibinfo{volume}{17}},
  \bibinfo{pages}{157} (\bibinfo{year}{1996}).

\bibitem[{\citenamefont{Ihm et~al.}(1997)\citenamefont{Ihm, Schneider, and
  Nielaba}}]{ISN97}
\bibinfo{author}{\bibfnamefont{M.-O.} \bibnamefont{Ihm}},
  \bibinfo{author}{\bibfnamefont{F.}~\bibnamefont{Schneider}},
  \bibnamefont{and} \bibinfo{author}{\bibfnamefont{P.}~\bibnamefont{Nielaba}},
  \bibinfo{journal}{Prog. Colloid Polym. Sci.} \textbf{\bibinfo{volume}{104}},
  \bibinfo{pages}{166} (\bibinfo{year}{1997}).

\bibitem[{\citenamefont{Nielaba}(1997)}]{N97}
\bibinfo{author}{\bibfnamefont{P.}~\bibnamefont{Nielaba}}, in
  \emph{\bibinfo{booktitle}{Ann. Rev. Com. Phys.}}, edited by
  \bibinfo{editor}{\bibfnamefont{D.}~\bibnamefont{Stauffer}}
  (\bibinfo{publisher}{World Scientific}, \bibinfo{address}{Singapore},
  \bibinfo{year}{1997}), pp. \bibinfo{pages}{137--200}.

\bibitem[{\citenamefont{Saija et~al.}(1998)\citenamefont{Saija, Fiumara, and
  Giaquinta}}]{SFG98}
\bibinfo{author}{\bibfnamefont{F.}~\bibnamefont{Saija}},
  \bibinfo{author}{\bibfnamefont{G.}~\bibnamefont{Fiumara}}, \bibnamefont{and}
  \bibinfo{author}{\bibfnamefont{P.~V.} \bibnamefont{Giaquinta}},
  \bibinfo{journal}{J. Chem. Phys.} \textbf{\bibinfo{volume}{108}},
  \bibinfo{pages}{9098} (\bibinfo{year}{1998}).

\bibitem[{\citenamefont{Al-Naafa et~al.}(1999)\citenamefont{Al-Naafa,
  El-Yakubu, and Hamad}}]{AEH99}
\bibinfo{author}{\bibfnamefont{M.}~\bibnamefont{Al-Naafa}},
  \bibinfo{author}{\bibfnamefont{J.~B.} \bibnamefont{El-Yakubu}},
  \bibnamefont{and} \bibinfo{author}{\bibfnamefont{E.~Z.} \bibnamefont{Hamad}},
  \bibinfo{journal}{Fluid Phase Equil.} \textbf{\bibinfo{volume}{154}},
  \bibinfo{pages}{33} (\bibinfo{year}{1999}).

\bibitem[{\citenamefont{Hamad and Yahaya}(2000)}]{HY00}
\bibinfo{author}{\bibfnamefont{E.~Z.} \bibnamefont{Hamad}} \bibnamefont{and}
  \bibinfo{author}{\bibfnamefont{G.~O.} \bibnamefont{Yahaya}},
  \bibinfo{journal}{Fluid Phase Equil.} \textbf{\bibinfo{volume}{168}},
  \bibinfo{pages}{59} (\bibinfo{year}{2000}).

\bibitem[{\citenamefont{Nielaba}(2000)}]{N00}
\bibinfo{author}{\bibfnamefont{P.}~\bibnamefont{Nielaba}}, in
  \emph{\bibinfo{booktitle}{Computational Methods in Surface and Colloid
  Science}}, edited by
  \bibinfo{editor}{\bibfnamefont{M.}~\bibnamefont{Bor\'owko}}
  (\bibinfo{publisher}{CRC Press}, \bibinfo{address}{Boca Raton},
  \bibinfo{year}{2000}), vol.~\bibinfo{volume}{89} of
  \emph{\bibinfo{series}{Surfactant Science Series}}, pp.
  \bibinfo{pages}{77--134}.

\bibitem[{\citenamefont{Saija and Giaquinta}(2002)}]{SG02}
\bibinfo{author}{\bibfnamefont{F.}~\bibnamefont{Saija}} \bibnamefont{and}
  \bibinfo{author}{\bibfnamefont{P.~V.} \bibnamefont{Giaquinta}},
  \bibinfo{journal}{J. Chem. Phys.} \textbf{\bibinfo{volume}{117}},
  \bibinfo{pages}{5780} (\bibinfo{year}{2002}).

\bibitem[{\citenamefont{Casta{\~n}eda-Priego
  et~al.}(2003)\citenamefont{Casta{\~n}eda-Priego, Rodr\'{\i}guez-L\'opez, and
  Alcaraz}}]{CRM03}
\bibinfo{author}{\bibfnamefont{R.}~\bibnamefont{Casta{\~n}eda-Priego}},
  \bibinfo{author}{\bibfnamefont{A.}~\bibnamefont{Rodr\'{\i}guez-L\'opez}},
  \bibnamefont{and} \bibinfo{author}{\bibfnamefont{J.~M.~M.}
  \bibnamefont{Alcaraz}}, \bibinfo{journal}{J. Phys.: Condens. Matter}
  \textbf{\bibinfo{volume}{15}}, \bibinfo{pages}{S3393} (\bibinfo{year}{2003}).

\bibitem[{\citenamefont{Faller and Kuhl}(2003)}]{FK03}
\bibinfo{author}{\bibfnamefont{R.}~\bibnamefont{Faller}} \bibnamefont{and}
  \bibinfo{author}{\bibfnamefont{T.~L.} \bibnamefont{Kuhl}},
  \bibinfo{journal}{Soft Matter.} \textbf{\bibinfo{volume}{1}},
  \bibinfo{pages}{343} (\bibinfo{year}{2003}).

\bibitem[{\citenamefont{Santos et~al.}(2005)\citenamefont{Santos, {L\'opez de
  Haro}, and Yuste}}]{SHY05}
\bibinfo{author}{\bibfnamefont{A.}~\bibnamefont{Santos}},
  \bibinfo{author}{\bibfnamefont{M.}~\bibnamefont{{L\'opez de Haro}}},
  \bibnamefont{and} \bibinfo{author}{\bibfnamefont{S.~B.} \bibnamefont{Yuste}},
  \bibinfo{journal}{J. Chem. Phys.} \textbf{\bibinfo{volume}{122}},
  \bibinfo{pages}{{024}{514}} (\bibinfo{year}{2005}).

\bibitem[{\citenamefont{Buhot}(2005)}]{B05}
\bibinfo{author}{\bibfnamefont{A.}~\bibnamefont{Buhot}}, \bibinfo{journal}{J.
  Chem. Phys.} \textbf{\bibinfo{volume}{122}}, \bibinfo{pages}{024105}
  (\bibinfo{year}{2005}).

\bibitem[{\citenamefont{Hoffmann et~al.}(2006)\citenamefont{Hoffmann, Likos,
  and L\"owen}}]{HLL06}
\bibinfo{author}{\bibfnamefont{N.}~\bibnamefont{Hoffmann}},
  \bibinfo{author}{\bibfnamefont{C.~N.} \bibnamefont{Likos}}, \bibnamefont{and}
  \bibinfo{author}{\bibfnamefont{H.}~\bibnamefont{L\"owen}},
  \bibinfo{journal}{J. Phys.: Condens. Matter} \textbf{\bibinfo{volume}{18}},
  \bibinfo{pages}{10193} (\bibinfo{year}{2006}).

\bibitem[{\citenamefont{Barcenas et~al.}(2008)\citenamefont{Barcenas, Orea,
  Buenrostro-Gonz\'alez, Zamudio-Rivera, and Duda}}]{BOBZD08}
\bibinfo{author}{\bibfnamefont{M.}~\bibnamefont{Barcenas}},
  \bibinfo{author}{\bibfnamefont{P.}~\bibnamefont{Orea}},
  \bibinfo{author}{\bibfnamefont{E.}~\bibnamefont{Buenrostro-Gonz\'alez}},
  \bibinfo{author}{\bibfnamefont{L.~S.} \bibnamefont{Zamudio-Rivera}},
  \bibnamefont{and} \bibinfo{author}{\bibfnamefont{Y.}~\bibnamefont{Duda}},
  \bibinfo{journal}{Energy \& Fuels} \textbf{\bibinfo{volume}{22}},
  \bibinfo{pages}{1917} (\bibinfo{year}{2008}).

\bibitem[{\citenamefont{Gu\'aqueta}(2009)}]{G09}
\bibinfo{author}{\bibfnamefont{R.~C.} \bibnamefont{Gu\'aqueta}}, Ph.D. thesis,
  \bibinfo{school}{University of Illinois, Urbana-Champaign}
  (\bibinfo{year}{2009}).

\bibitem[{\citenamefont{Mu{\~n}oz-Salazar and Odriozola}(2010)}]{MO10}
\bibinfo{author}{\bibfnamefont{L.}~\bibnamefont{Mu{\~n}oz-Salazar}}
  \bibnamefont{and}
  \bibinfo{author}{\bibfnamefont{G.}~\bibnamefont{Odriozola}},
  \bibinfo{journal}{Mol. Simul.} \textbf{\bibinfo{volume}{36}},
  \bibinfo{pages}{175} (\bibinfo{year}{2010}).

\bibitem[{\citenamefont{Santos et~al.}(2010)\citenamefont{Santos, {L\'opez de
  Haro}, and Yuste}}]{SHY10}
\bibinfo{author}{\bibfnamefont{A.}~\bibnamefont{Santos}},
  \bibinfo{author}{\bibfnamefont{M.}~\bibnamefont{{L\'opez de Haro}}},
  \bibnamefont{and} \bibinfo{author}{\bibfnamefont{S.~B.} \bibnamefont{Yuste}},
  \bibinfo{journal}{J. Chem. Phys.} \textbf{\bibinfo{volume}{132}},
  \bibinfo{pages}{204506} (\bibinfo{year}{2010}).

\bibitem[{\citenamefont{Saija}(2011)}]{S11}
\bibinfo{author}{\bibfnamefont{F.}~\bibnamefont{Saija}},
  \bibinfo{journal}{Phys. Chem. Chem. Phys.} \textbf{\bibinfo{volume}{13}},
  \bibinfo{pages}{{118}{85}} (\bibinfo{year}{2011}).

\bibitem[{\citenamefont{Saija et~al.}(2012)\citenamefont{Saija, Santos, Yuste,
  and {L\'opez de Haro}}}]{SSYH12}
\bibinfo{author}{\bibfnamefont{F.}~\bibnamefont{Saija}},
  \bibinfo{author}{\bibfnamefont{A.}~\bibnamefont{Santos}},
  \bibinfo{author}{\bibfnamefont{S.~B.} \bibnamefont{Yuste}}, \bibnamefont{and}
  \bibinfo{author}{\bibfnamefont{M.}~\bibnamefont{{L\'opez de Haro}}},
  \bibinfo{journal}{J. Chem. Phys.} \textbf{\bibinfo{volume}{136}},
  \bibinfo{pages}{{184}{505}} (\bibinfo{year}{2012}).

\bibitem[{\citenamefont{Fiumara et~al.}(2014)\citenamefont{Fiumara, Pandaram,
  Pellicane, and Saija}}]{FPPS14}
\bibinfo{author}{\bibfnamefont{G.}~\bibnamefont{Fiumara}},
  \bibinfo{author}{\bibfnamefont{O.~D.} \bibnamefont{Pandaram}},
  \bibinfo{author}{\bibfnamefont{G.}~\bibnamefont{Pellicane}},
  \bibnamefont{and} \bibinfo{author}{\bibfnamefont{F.}~\bibnamefont{Saija}},
  \bibinfo{journal}{J. Chem. Phys.} \textbf{\bibinfo{volume}{141}},
  \bibinfo{pages}{214508} (\bibinfo{year}{2014}).

\bibitem[{\citenamefont{G\'o\'zd\'z and Ciach}(2016)}]{GC16}
\bibinfo{author}{\bibfnamefont{W.~T.} \bibnamefont{G\'o\'zd\'z}}
  \bibnamefont{and} \bibinfo{author}{\bibfnamefont{A.}~\bibnamefont{Ciach}},
  \bibinfo{journal}{Condens. Matter Phys.} \textbf{\bibinfo{volume}{19}},
  \bibinfo{pages}{13002} (\bibinfo{year}{2016}).

\bibitem[{\citenamefont{Santos}(2016)}]{S16}
\bibinfo{author}{\bibfnamefont{A.}~\bibnamefont{Santos}},
  \emph{\bibinfo{title}{{A Concise Course on the Theory of Classical Liquids.
  Basics and Selected Topics}}}, vol. \bibinfo{volume}{923} of
  \emph{\bibinfo{series}{Lecture Notes in Physics}}
  (\bibinfo{publisher}{Springer}, \bibinfo{address}{New York},
  \bibinfo{year}{2016}).

\bibitem[{\citenamefont{Lab\'ik et~al.}(2005)\citenamefont{Lab\'ik, Kolafa, and
  Malijevsk\'y}}]{LKM05}
\bibinfo{author}{\bibfnamefont{S.}~\bibnamefont{Lab\'ik}},
  \bibinfo{author}{\bibfnamefont{J.}~\bibnamefont{Kolafa}}, \bibnamefont{and}
  \bibinfo{author}{\bibfnamefont{A.}~\bibnamefont{Malijevsk\'y}},
  \bibinfo{journal}{Phys. Rev. E} \textbf{\bibinfo{volume}{71}},
  \bibinfo{pages}{{021}{105}} (\bibinfo{year}{2005}).

\bibitem[{\citenamefont{Hamad}(1994)}]{H94}
\bibinfo{author}{\bibfnamefont{E.}~\bibnamefont{Hamad}}, \bibinfo{journal}{J.
  Chem. Phys.} \textbf{\bibinfo{volume}{101}}, \bibinfo{pages}{10195}
  (\bibinfo{year}{1994}).

\bibitem[{\citenamefont{Hamad}(1996{\natexlab{a}})}]{H96}
\bibinfo{author}{\bibfnamefont{E.~Z.} \bibnamefont{Hamad}},
  \bibinfo{journal}{J. Chem. Phys.} \textbf{\bibinfo{volume}{105}},
  \bibinfo{pages}{3222} (\bibinfo{year}{1996}{\natexlab{a}}).

\bibitem[{\citenamefont{Hamad}(1996{\natexlab{b}})}]{H96b}
\bibinfo{author}{\bibfnamefont{E.~Z.} \bibnamefont{Hamad}},
  \bibinfo{journal}{J. Chem. Phys.} \textbf{\bibinfo{volume}{105}},
  \bibinfo{pages}{3229} (\bibinfo{year}{1996}{\natexlab{b}}).

\bibitem[{\citenamefont{Baus and Colot}(1987)}]{BC87}
\bibinfo{author}{\bibfnamefont{M.}~\bibnamefont{Baus}} \bibnamefont{and}
  \bibinfo{author}{\bibfnamefont{J.~L.} \bibnamefont{Colot}},
  \bibinfo{journal}{Phys. Rev. A} \textbf{\bibinfo{volume}{36}},
  \bibinfo{pages}{3912} (\bibinfo{year}{1987}).

\bibitem[{\citenamefont{Barrat et~al.}(1988)\citenamefont{Barrat, Xu, Hansen,
  and Baus}}]{BXHB88}
\bibinfo{author}{\bibfnamefont{J.-L.} \bibnamefont{Barrat}},
  \bibinfo{author}{\bibfnamefont{H.}~\bibnamefont{Xu}},
  \bibinfo{author}{\bibfnamefont{J.-P.} \bibnamefont{Hansen}},
  \bibnamefont{and} \bibinfo{author}{\bibfnamefont{M.}~\bibnamefont{Baus}},
  \bibinfo{journal}{J. Phys. C} \textbf{\bibinfo{volume}{21}},
  \bibinfo{pages}{3165} (\bibinfo{year}{1988}).

\bibitem[{\citenamefont{Saija et~al.}(1996)\citenamefont{Saija, Fiumara, and
  Giaquinta}}]{SFG96}
\bibinfo{author}{\bibfnamefont{F.}~\bibnamefont{Saija}},
  \bibinfo{author}{\bibfnamefont{G.}~\bibnamefont{Fiumara}}, \bibnamefont{and}
  \bibinfo{author}{\bibfnamefont{P.~V.} \bibnamefont{Giaquinta}},
  \bibinfo{journal}{Mol. Phys.} \textbf{\bibinfo{volume}{87}},
  \bibinfo{pages}{991} (\bibinfo{year}{1996}), \bibinfo{note}{erratum:
  \textbf{92}, 1089 (1997)}.

\bibitem[{\citenamefont{Wheatley}(2013)}]{W13}
\bibinfo{author}{\bibfnamefont{R.~J.} \bibnamefont{Wheatley}},
  \bibinfo{journal}{Phys. Rev. Lett.} \textbf{\bibinfo{volume}{110}},
  \bibinfo{pages}{{200}{601}} (\bibinfo{year}{2013}).

\bibitem[{\citenamefont{Zhang and Pettitt}(2014)}]{ZP14}
\bibinfo{author}{\bibfnamefont{C.}~\bibnamefont{Zhang}} \bibnamefont{and}
  \bibinfo{author}{\bibfnamefont{B.~M.} \bibnamefont{Pettitt}},
  \bibinfo{journal}{Mol. Phys.} \textbf{\bibinfo{volume}{112}},
  \bibinfo{pages}{1427} (\bibinfo{year}{2014}).

\bibitem[{\citenamefont{Boubl{\'\i }k and Nezbeda}(1986)}]{BN86}
\bibinfo{author}{\bibfnamefont{T.}~\bibnamefont{Boubl{\'\i }k}}
  \bibnamefont{and} \bibinfo{author}{\bibfnamefont{I.}~\bibnamefont{Nezbeda}},
  \bibinfo{journal}{Coll. Czech. Chem. Commun.} \textbf{\bibinfo{volume}{51}},
  \bibinfo{pages}{2301} (\bibinfo{year}{1986}).

\bibitem[{\citenamefont{Bor{\u s}tnik}(1992)}]{B92}
\bibinfo{author}{\bibfnamefont{B.}~\bibnamefont{Bor{\u s}tnik}},
  \bibinfo{journal}{Vestn. Slov. Kem. Drus.} \textbf{\bibinfo{volume}{39}},
  \bibinfo{pages}{145} (\bibinfo{year}{1992}).

\bibitem[{\citenamefont{Kratky}(1977)}]{K77}
\bibinfo{author}{\bibfnamefont{K.~W.} \bibnamefont{Kratky}},
  \bibinfo{journal}{Physica A} \textbf{\bibinfo{volume}{87}},
  \bibinfo{pages}{584} (\bibinfo{year}{1977}).

\bibitem[{\citenamefont{Wheatley}(1998)}]{W98c}
\bibinfo{author}{\bibfnamefont{R.~J.} \bibnamefont{Wheatley}},
  \bibinfo{journal}{Mol. Phys.} \textbf{\bibinfo{volume}{93}},
  \bibinfo{pages}{665} (\bibinfo{year}{1998}).

\bibitem[{\citenamefont{Luding}(2001)}]{L01b}
\bibinfo{author}{\bibfnamefont{S.}~\bibnamefont{Luding}},
  \bibinfo{journal}{Phys. Rev. E} \textbf{\bibinfo{volume}{63}},
  \bibinfo{pages}{{042}{201}} (\bibinfo{year}{2001}).

\bibitem[{\citenamefont{Luding and Strau\ss}(2001)}]{LS01}
\bibinfo{author}{\bibfnamefont{S.}~\bibnamefont{Luding}} \bibnamefont{and}
  \bibinfo{author}{\bibfnamefont{O.}~\bibnamefont{Strau\ss}}, in
  \emph{\bibinfo{booktitle}{{Granular Gases}}}, edited by
  \bibinfo{editor}{\bibfnamefont{T.}~\bibnamefont{P{\"o}schel}}
  \bibnamefont{and} \bibinfo{editor}{\bibfnamefont{S.}~\bibnamefont{Luding}}
  (\bibinfo{publisher}{Springer}, \bibinfo{address}{Berlin},
  \bibinfo{year}{2001}), vol. \bibinfo{volume}{564} of
  \emph{\bibinfo{series}{Lecture Notes in Physics}}, pp.
  \bibinfo{pages}{389--409}.

\bibitem[{\citenamefont{Luding and Santos}(2004)}]{LS04}
\bibinfo{author}{\bibfnamefont{S.}~\bibnamefont{Luding}} \bibnamefont{and}
  \bibinfo{author}{\bibfnamefont{A.}~\bibnamefont{Santos}},
  \bibinfo{journal}{J. Chem. Phys.} \textbf{\bibinfo{volume}{121}},
  \bibinfo{pages}{8458} (\bibinfo{year}{2004}).

\bibitem[{\citenamefont{Panagiotopoulos}(1992)}]{P92}
\bibinfo{author}{\bibfnamefont{A.~Z.} \bibnamefont{Panagiotopoulos}},
  \bibinfo{journal}{Mol. Simul.} \textbf{\bibinfo{volume}{9}},
  \bibinfo{pages}{1} (\bibinfo{year}{1992}).

\bibitem[{\citenamefont{Frenkel and Smit}(2002)}]{FS02}
\bibinfo{author}{\bibfnamefont{D.}~\bibnamefont{Frenkel}} \bibnamefont{and}
  \bibinfo{author}{\bibfnamefont{B.}~\bibnamefont{Smit}},
  \emph{\bibinfo{title}{Understanding Molecular Simulation: From Algorithms to
  Applications}} (\bibinfo{publisher}{Academic Press}, \bibinfo{address}{San
  Diego}, \bibinfo{year}{2002}), \bibinfo{edition}{2nd} ed.

\bibitem[{\citenamefont{Widom}(1963)}]{W63b}
\bibinfo{author}{\bibfnamefont{B.}~\bibnamefont{Widom}}, \bibinfo{journal}{J.
  Chem. Phys.} \textbf{\bibinfo{volume}{39}}, \bibinfo{pages}{2808}
  (\bibinfo{year}{1963}).

\end{thebibliography}

\setcounter{table}{0}



\appendix

\section*{Supplementary material to the paper ``Virial coefficients, equation of state, and demixing of  binary asymmetric nonadditive  hard-disk mixtures'' }

The following tables contain the results of our computations of the composition-independent fifth virial coefficients (Tables \ref{p005quinto}--\ref{p06quinto}), compressibility factor (Tables \ref{q04_D03} and \ref{q05_D03}), and binodal curves (Table \ref{Demix_q04_D03}) of asymmetric nonadditive hard-disk mixtures.


\begin{table*}
\caption{Fifth-order virial coefficients $E_{11112}$, $E_{11122}$, $E_{11222}$, and $E_{12222}$ as  functions of the size ratio
$q=\sigma_2/\sigma_1$ for $\Delta = 0.05$. The error on the last significant figure is enclosed between parentheses.}
\label{p005quinto}
\begin{ruledtabular}
\begin{tabular}{ddddd}
\multicolumn{1}{c}{$\quad q$}  & \multicolumn{1}{c}{$\qquad\quad E_{11112}/\sigma_1^8$} & \multicolumn{1}{c}{$\qquad\qquad E_{11122}/\sigma_1^8$} & \multicolumn{1}{c}{$\qquad\qquad E_{11222}/\sigma_1^8$} &\multicolumn{1}{c}{$\qquad\qquad E_{12222}/\sigma_1^8$} \\
\hline
 0.10    &  0.2070(2) &  5.35(3)\times 10^{-3}    & 9.83(1)\times 10^{-5}    & 1.511(3)\times 10^{-6}  \\
 0.20    &  0.3329(2) &  2.7622(5)\times 10^{-2}  & 1.801(4)\times 10^{-3}   & 1.022(1)\times 10^{-4}  \\
 0.30    &  0.4874(3) &  7.751(2)\times 10^{-2}   & 1.032(3)\times 10^{-2}   & 1.228(2)\times 10^{-3}  \\
 0.40    &  0.6708(9) &  0.1677(3)                & 3.658(6)\times 10^{-2}   & 7.27(1)\times 10^{-3}   \\
 0.50    &  0.883(1)  &  0.3131(5)                & 9.95(1)\times 10^{-2}    & 2.920(4)\times 10^{-2}  \\
 0.60    &  1.123(2)  &  0.5306(9)                & 0.2285(4)                & 9.16(2)\times 10^{-2}   \\
 0.66667 &  1.299(2)  &  0.725(1)                 & 0.3719(9)                & 0.1781(3)               \\
 0.70    &  1.392(2)  &  0.839(1)                 & 0.4664(7)                & 0.2425(4)               \\
 0.80    &  1.689(3)  &  1.259(3)                 & 0.873(2)                 & 0.567(1)                \\
 0.85    &  1.848(2)  &  1.518(2)                 & 1.163(1)                 & 0.834(1)                \\
 0.90    &  2.014(4)  &  1.814(4)                 & 1.526(3)                 & 1.203(2)                \\
 0.95    &  2.188(3)  &  2.150(3)               & 1.976(3)                 & 1.702(2)              \\
 1.00    & 2.368(2)  & 2.529(2)              & 2.529(2)                & 2.368(2)              \\
\end {tabular}
\end{ruledtabular}
\end{table*}


\begin{table*}
\caption{Same as in Table \ref{p005quinto}, but  for $\Delta = 0.1$.}
\label{p01quinto}
\begin{ruledtabular}
\begin{tabular}{ddddd}
\multicolumn{1}{c}{$\quad q$}  & \multicolumn{1}{c}{$\qquad\quad E_{11112}/\sigma_1^8$} & \multicolumn{1}{c}{$\qquad\qquad E_{11122}/\sigma_1^8$} & \multicolumn{1}{c}{$\qquad\qquad E_{11222}/\sigma_1^8$} &\multicolumn{1}{c}{$\qquad\qquad E_{12222}/\sigma_1^8$} \\
\hline
  0.10    &  0.2693(2) &  6.40(3)\times 10^{-3}    & 1.09(2)\times 10^{-4}   & 1.669(3)\times 10^{-6}  \\
 0.20    &  0.4177(3) &  3.3538(6)\times 10^{-2}  & 2.083(6)\times 10^{-3}  &  1.135(1)\times 10^{-4}  \\
 0.30    &  0.5977(8) &  9.41(3)\times 10^{-2}    & 1.208(4)\times 10^{-2}   & 1.372(3)\times 10^{-3}  \\
 0.40    &  0.809(1)  &  0.2035(4)                & 4.311(8)\times 10^{-2}   & 8.17(1)\times 10^{-3}   \\
 0.50    &  1.051(1)  &  0.3790(6)                & 0.1178(2)                & 3.295(5)\times 10^{-2}  \\
 0.60    &  1.325(2)  &  0.641(1)                 & 0.2714(5)                & 0.1039(2)               \\
 0.66667 &  1.525(2)  &  0.874(1)                 & 0.4424(7)                & 0.2026(3)               \\
 0.70    &  1.630(2)  &  1.011(2)                 & 0.5554(9)                & 0.2763(5)               \\
 0.80    &  1.965(4)  &  1.514(4)                 & 1.041(2)                 & 0.648(1)                \\
 0.85    &  2.1459(8) &  1.8240(8)                & 1.3892(6)                & 0.9569(4)               \\
 0.90    &  2.332(4)  &  2.177(5)                 & 1.825(4)                 & 1.382(3)                \\
 0.95    &  2.529(1) & 2.579(1)               & 2.366(1)                 & 1.9600(8)               \\
  1.00    & 2.730(2)  & 3.030(2)             & 3.030(2)                &2.730(2)              \\
\end {tabular}
\end{ruledtabular}
\end{table*}


\begin{table*}
\caption{Same as in Table \ref{p005quinto}, but  for $\Delta = 0.15$.}
\label{p015quinto}
\begin{ruledtabular}
\begin{tabular}{ddddd}
\multicolumn{1}{c}{$\quad q$}  & \multicolumn{1}{c}{$\qquad\quad E_{11112}/\sigma_1^8$} & \multicolumn{1}{c}{$\qquad\qquad E_{11122}/\sigma_1^8$} & \multicolumn{1}{c}{$\qquad\qquad E_{11222}/\sigma_1^8$} &\multicolumn{1}{c}{$\qquad\qquad E_{12222}/\sigma_1^8$} \\
\hline
  0.10    &  0.3395(2) &  6.87(5)\times 10^{-3}    & 9.98(2)\times 10^{-5}    & 1.835(4)\times 10^{-6}  \\
 0.20    &  0.5120(3) &  3.824(8)\times 10^{-2}   & 2.249(8)\times 10^{-3}   & 1.255(1)\times 10^{-4}  \\
 0.30    &  0.7188(8) &  0.1087(3)                & 1.344(5)\times 10^{-2}   & 1.524(3)\times 10^{-3}  \\
 0.40    &  0.960(1)  &  0.2357(5)                & 4.861(8)\times 10^{-2}   & 9.11(2)\times 10^{-3}   \\
 0.50    &  1.235(1)  &  0.4392(8)                & 0.1338(2)                & 3.694(6)\times 10^{-2}  \\
 0.60    &  1.544(2)  &  0.742(2)                 & 0.3099(6)                & 0.1170(2)               \\
 0.66667 &  1.768(2)  &  1.012(2)                 & 0.5063(1)                & 0.2287(4)               \\
 0.70    &  1.887(3)  &  1.170(2)                 & 0.637(1)                 & 0.3124(5)               \\
 0.80    &  2.263(4)  &  1.750(5)                 & 1.196(3)                 & 0.735(2)                \\
 0.85    &  2.465(3)  &  2.107(3)                 & 1.597(3)                 & 1.087(2)               \\
 0.90    &  2.674(5)  &  2.514(6)                 & 2.100(5)                 & 1.573(3)                \\
 0.95    &  2.894(4)  &  2.975(5)                & 2.726(4)               & 2.235(3)           \\
  1.00    & 3.120(2)  & 3.494(3)             & 3.494(3)                &3.120(2)             \\
\end {tabular}
\end{ruledtabular}
\end{table*}


\begin{table*}
\caption{Same as in Table \ref{p005quinto}, but  for $\Delta = 0.2$.}
\label{p02quinto}
\begin{ruledtabular}
\begin{tabular}{ddddd}
\multicolumn{1}{c}{$\quad q$}  & \multicolumn{1}{c}{$\qquad\quad E_{11112}/\sigma_1^8$} & \multicolumn{1}{c}{$\qquad\qquad E_{11122}/\sigma_1^8$} & \multicolumn{1}{c}{$\qquad\qquad E_{11222}/\sigma_1^8$} &\multicolumn{1}{c}{$\qquad\qquad E_{12222}/\sigma_1^8$} \\
\hline
 0.10    &  0.4177(3) &  6.12(6)\times 10^{-3}    & 4.32(3)\times 10^{-5}    & 2.009(5)\times 10^{-6}  \\
 0.20    &  0.6157(3) &  4.03(1)\times 10^{-2}    & 2.17(1)\times 10^{-3}    & 1.380(2)\times 10^{-4}  \\
 0.30    &  0.851(1)  &  0.1182(4)                & 1.397(7)\times 10^{-2}   & 1.684(4)\times 10^{-3}  \\
 0.40    &  1.123(1)  &  0.2594(7)                & 5.19(1)\times 10^{-2}    & 1.011(2)\times 10^{-2}   \\
 0.50    &  1.433(1)  &  0.486(1)                 & 0.1448(3)                & 4.115(6)\times 10^{-2}  \\
 0.60    &  1.779(2)  &  0.822(2)                 & 0.3384(8)                & 0.1308(3)               \\
 0.66667 &  2.030(2)  &  1.122(2)                 & 0.555(1)                 & 0.2564(4)               \\
 0.70    &  2.162(3)  &  1.297(3)                 & 0.698(2)                 & 0.3504(3)               \\
 0.80    &  2.583(5)  &  1.939(6)                 & 1.317(4)                 & 0.828(2)                \\
 0.85    &  2.808(1)  &  2.334(1)                 & 1.761(1)                 & 1.2266(6)               \\
 0.90    &  3.040(6)  &  2.782(8)                 & 2.319(7)                 & 1.777(4)                \\
 0.95    &  3.285(1)  &  3.292(2)                & 3.012(2)                & 2.529(1)                \\
 1.00    &  3.535(2)  & 3.863(3)               & 3.863(3)                &3.535(2)                \\
\end {tabular}
\end{ruledtabular}
\end{table*}


\begin{table*}
\caption{Same as in Table \ref{p005quinto}, but  for $\Delta = 0.3$.}
\label{p03quinto}
\begin{ruledtabular}
\begin{tabular}{ddddd}
\multicolumn{1}{c}{$\quad q$}  & \multicolumn{1}{c}{$\qquad\quad E_{11112}/\sigma_1^8$} & \multicolumn{1}{c}{$\qquad\qquad E_{11122}/\sigma_1^8$} & \multicolumn{1}{c}{$\qquad\qquad E_{11222}/\sigma_1^8$} &\multicolumn{1}{c}{$\qquad\qquad E_{12222}/\sigma_1^8$} \\
\hline
 0.10    &  0.5977(3) & -3.03(6)\times 10^{-3}    & -4.27(5)\times 10^{-4}   & 2.378(5)\times 10^{-6}  \\
 0.20    &  0.8508(4) &  2.79(2)\times 10^{-2}    &  3.51(2)\times 10^{-4}   & 1.649(2)\times 10^{-4}  \\
 0.30    &  1.148(1)  &  0.1067(7)                &  9.72(7)\times 10^{-3}   & 2.027(5)\times 10^{-3}  \\
 0.40    &  1.488(2)  &  0.254(1)                 &  4.47(3)\times 10^{-2}   & 1.226(2)\times 10^{-2}   \\
 0.50    &  1.871(2)  &  0.493(2)                 &  0.1360(5)               & 5.026(9)\times 10^{-2}  \\
 0.60    &  2.299(3)  &  0.849(3)                 &  0.332(1)                & 0.1608(3)               \\
 0.66667 &  2.607(3)  &  1.164(4)                 &  0.554(1)                & 0.3164(6)               \\
 0.70    &  2.770(4)  &  1.349(5)                 &  0.702(3)                & 0.4333(8)               \\
 0.80    &  3.284(7)  &  2.021(9)                 &  1.345(7)                & 1.030(3)                \\
 0.85    &  3.558(2)  &  2.433(2)                 &  1.809(1)                & 1.5297(7)               \\
 0.90    &  3.843(8)  &  2.90(1)                  &  2.39(1)                 & 2.223(5)                \\
 0.95    &  4.138(2)  &  3.422(3)                  &  3.117(3)                 & 3.171(1)                \\
 1.00    &  4.444(2)  & 4.007(5)              & 4.007(5)               &4.444(2)                \\
\end {tabular}
\end{ruledtabular}
\end{table*}


\begin{table*}
\caption{Same as in Table \ref{p005quinto}, but  for $\Delta = 0.4$.}
\label{p04quinto}
\begin{ruledtabular}
\begin{tabular}{ddddd}
\multicolumn{1}{c}{$\quad q$}  & \multicolumn{1}{c}{$\qquad\quad E_{11112}/\sigma_1^8$} & \multicolumn{1}{c}{$\qquad\qquad E_{11122}/\sigma_1^8$} & \multicolumn{1}{c}{$\qquad\qquad E_{11222}/\sigma_1^8$} &\multicolumn{1}{c}{$\qquad\qquad E_{12222}/\sigma_1^8$} \\
\hline
 0.10    &  0.8088(3) & -3.25(4)\times 10^{-2}    & -2.100(9)\times 10^{-3}  & 2.780(7)\times 10^{-6}  \\
 0.20    &  1.1229(5) & -2.73(3)\times 10^{-2}    & -6.30(3)\times 10^{-3}         & 1.940(3)\times 10^{-4}\\
 0.30    &  1.487(2)  &  1.52(1)\times 10^{-2}    & -9.22(8)\times 10^{-3}         & 2.404(6)\times 10^{-3}\\
 0.40    &  1.902(2)  &  0.111(2)                 &  4.5(5)\times 10^{-4}          & 1.464(3)\times 10^{-2}\\
 0.50    &  2.368(2)  &  0.275(3)                 &  4.451(1)\times 10^{-2}        & 6.03(4)\times 10^{-2}\\
 0.60    &  2.884(4)  &  0.525(5)                 &  0.15812(6)                    & 0.1940(4)    \\
 0.66667 &  3.256(4)  &  0.745(6)                 &  0.297(3)                      & 0.3827(7) \\
 0.70    &  3.451(5)  &  0.873(8)                 &  0.392(1)                      & 0.525(1)      \\
 0.80    &  4.071(9)  &  1.33(2)                  &  0.81(1)                       & 1.255(3)   \\
 0.85    &  4.397(2)  &  1.605(4)                 &  1.125(2)                      & 1.866(6) \\
0.90    &  4.74(2)   &  1.91(2)                  &  1.52(2)                       & 2.719(7) \\
0.95    &  5.091(2)   &  2.242(5)                  &  2.006(5)                       & 3.884(2) \\
1.00    &  5.458(2)   & 2.60(1)                 &  2.60(1)                 & 5.458(2) \\
\end {tabular}
\end{ruledtabular}
\end{table*}


\begin{table*}
\caption{Same as in Table \ref{p005quinto}, but  for $\Delta = 0.5$.}
\label{p05quinto}
\begin{ruledtabular}
\begin{tabular}{ddddd}
\multicolumn{1}{c}{$\quad q$}  & \multicolumn{1}{c}{$\qquad\quad E_{11112}/\sigma_1^8$} & \multicolumn{1}{c}{$\qquad\qquad E_{11122}/\sigma_1^8$} & \multicolumn{1}{c}{$\qquad\qquad E_{11222}/\sigma_1^8$} &\multicolumn{1}{c}{$\qquad\qquad E_{12222}/\sigma_1^8$} \\
\hline
 0.10    &  1.0516(5) & -0.1008(5)                & -6.66(1)\times 10^{-3}   & 3.214(9)\times 10^{-6}  \\
 0.20    &  1.4324(6) & -0.1627(4)                & -2.326(5)\times 10^{-2}  & 2.258(3)\times 10^{-4}  \\
 0.30    &  1.871(2)  & -0.224(2)                 & -5.77(3)\times 10^{-2}   & 2.811(8)\times 10^{-3}  \\
 0.40    &  2.368(2)  & -0.286(3)                 & -0.1162(7)               & 1.720(4)\times 10^{-2}   \\
 0.50    &  2.923(3)  & -0.354(4)                 & -0.205(1)                & 7.12(1)\times 10^{-2}  \\
 0.60    &  3.535(5)  & -0.444(8)                 & -0.331(3)                & 0.2301(6)               \\
 0.66667 &  3.977(6)  & -0.525(9)                 & -0.442(4)                & 0.4554(9)               \\
 0.70    &  4.207(7)  & -0.58(1)                  & -0.508(6)                & 0.626(1)               \\
 0.80    &  4.94(1)   & -0.79(2)                  & -0.76(2)                 & 1.500(4)                \\
 0.85    &  5.32(1)   & -0.94(1)                  & -0.92(1)                 & 2.236(1)               \\
 0.90    &  5.72(1)   & -1.13(3)                  & -1.12(3)                 & 3.262(9)                \\
 0.95    &  6.141(3)   & -1.370(8)                  & -1.367(7)                 & 4.670(3)                \\
 1.00    &  6.569(3)   &-1.67(2)                 &  -1.67(2)                &6.569(3) \\
\end {tabular}
\end{ruledtabular}
\end{table*}


\begin{table*}
\caption{Same as in Table \ref{p005quinto}, but  for $\Delta = 0.6$.}
\label{p06quinto}
\begin{ruledtabular}
\begin{tabular}{ddddd}
\multicolumn{1}{c}{$\quad q$}  & \multicolumn{1}{c}{$\qquad\quad E_{11112}/\sigma_1^8$} & \multicolumn{1}{c}{$\qquad\qquad E_{11122}/\sigma_1^8$} & \multicolumn{1}{c}{$\qquad\qquad E_{11222}/\sigma_1^8$} &\multicolumn{1}{c}{$\qquad\qquad E_{12222}/\sigma_1^8$} \\
\hline
 0.10    &  1.3252(6) & -0.2358(2)                & -1.727(1)\times 10^{-2}  & 3.68(1)\times 10^{-6}  \\
 0.20    &  1.7788(7) & -0.4331(9)                & -5.997(8)\times 10^{-2}  & 2.598(4)\times 10^{-4}  \\
 0.30    &  2.2988(9) & -0.711(1)                 & -0.1601(2)               & 3.251(4)\times 10^{-3}  \\
 0.40    &  2.885(1)  & -1.106(2)                 & -0.3609(4)               & 1.998(2)\times 10^{-2}   \\
 0.50    &  3.537(2)  & -1.673(2)                 & -0.7281(8)               & 8.305(7)\times 10^{-2}  \\
 0.60    &  4.255(2)  & -2.486(3)                 & -1.362(1)                & 0.2693(2)               \\
 0.66667 &  4.770(7)  & -3.22(1)                  & -2.003(6)                & 0.534(1)               \\
 0.70    &  5.039(2)  & -3.650(5)                 & -2.411(3)                & 0.7351(5)               \\
 0.80    &  5.890(3)  & -5.296(7)                 & -4.097(5)                & 1.768(1)                \\
 0.85    &  6.34(1)   & -6.35(3)                  & -5.27(2)                 & 2.639(5)               \\
 0.90    &  6.807(4)  & -7.59(1)                  & -6.733(2)                & 3.857(2)                \\
 0.95    &  7.29(1)  & -9.06(4)                  & -8.54(3)                & 5.53(1)                \\
  1.00    & 7.789(3)   &-10.778(6)               & -10.778(6)            &7.789(3) \\
\end {tabular}
\end{ruledtabular}
\end{table*}


\begin{table*}
\caption{Compressibility factor $Z=\beta p/\rho$ as a function of density for a nonadditivity parameter $\Delta = 0.3$, a size ratio
 $q=\sigma_2/\sigma_1 = 0.4$,     and  several values of the mole fraction $x_1$. The error on the last significant figure is enclosed between parentheses.}
\label{q04_D03}
\begin{ruledtabular}
\begin{tabular}{dddddd}
\multicolumn{1}{c}{$\quad \rho\sigma_1^2$}       &\multicolumn{1}{c}{$\qquad x_1=0.1$}  &\multicolumn{1}{c}{$\qquad x_1=0.3$}&\multicolumn{1}{c}{$\qquad x_1=0.5$}&\multicolumn{1}{c}{$\qquad x_1=0.7$}&\multicolumn{1}{c}{$\qquad x_1=0.9$}  \\
\hline
0.20     &   1.0974(2)    &  1.1846(1)  &   1.2652(1)&   1.3348(2)&  1.3913(2)   \\
0.30     &   1.1514(3)    &1.2964(3) & 1.4351(3)&1.5691(2) &1.6805(3) \\
0.40     &   1.2119(5)     &1.4259(4) &1.6483(4) &1.8708(3) & 2.0711(5)\\
0.50     &   1.2741(1)    &1.571(1) &1.90061(4) &2.259(4) &2.614(1) \\
0.55     &   1.3079(3)      &1.651(2) &2.0478(1) &2.4964(7) &2.965(2) \\
0.60     &   1.3428(4)       &1.7390(2) &2.2098(3) &2.769(3) &3.389(2) \\
0.65     &   1.3804(6)     &1.8307(7)  &2.389(3) &3.081(2) & 3.9048(5)\\
0.70     &   1.4177(3)      &1.9284(7)  &2.585(3) &3.4387(7) &4.537(8) \\
0.75     &   1.457(1)     &2.035(1)   &2.798(2) &3.850(3) &5.314(2) \\
0.80     &   1.4959(6)     &2.144(1)    &3.034(3) &4.315(8) &6.271(7) \\
0.85     &   1.5392(5)      &2.261(1)  &3.281(1) &4.82(1) &7.49(1) \\
0.90     &   1.5835(4)    &2.387(2)   &3.554(4) &5.391(9) &9.02(2) \\
0.95     &   1.6287(6)     &2.516(3)   &3.85(1) &6.00(3) &10.8(1) \\
1.00     &   1.6768(2)    &2.656(3)     &4.14(2) &6.76(4) &12.7(1) \\
\end {tabular}
\end{ruledtabular}
\end{table*}


\begin{table*}
\caption{Same as in Table \ref{q04_D03}, but  for $\Delta = 0.3$ and $q=\sigma_2/\sigma_1 = 0.5$. The error on the last significant figure is enclosed between parentheses.}
\label{q05_D03}
\begin{ruledtabular}
\begin{tabular}{dddddd}
\multicolumn{1}{c}{$\quad \rho\sigma_1^2$}       &\multicolumn{1}{c}{$\qquad x_1=0.1$}  &\multicolumn{1}{c}{$\qquad x_1=0.3$}&\multicolumn{1}{c}{$\qquad x_1=0.5$}&\multicolumn{1}{c}{$\qquad x_1=0.7$}&\multicolumn{1}{c}{$\qquad x_1=0.9$}  \\
\hline
  0.20     &   1.1323(2)  &1.2237(1) &1.3021(2) &1.3630(3) &1.4019(2)  \\
  0.30     &   1.2105(1)  &1.3646(2) &1.5056(3) & 1.6226(4) &1.7037(3)    \\
  0.40     &     1.2946(4)  &1.5283(4) &1.7562(4) & 1.9584(5) &2.1110(3)    \\
  0.50     &    1.3891(3)     &1.720(2) &2.063(2) &2.400(1) &2.683(3)  \\
  0.55     &   1.4405(7)     &1.827(1) & 2.243(2)&2.668(1)  &3.054(2)  \\
  0.60     &   1.495(2)      &1.943(1) &2.440(2) &2.982(4)&3.503(1)    \\
  0.65     &   1.5502(8)      &2.069(2) &2.6575(9) &3.338(3) &4.0442(5)   \\
  0.70     &    1.6111(6)     &2.199(4) &2.898(2) & 3.731(5)&4.707(3)   \\
  0.75     &    1.6732(7)     &2.343(1) &3.159(1) & 4.18(1)&5.51(1)   \\
  0.80     &     1.7392(6)    &2.4940(8) &3.433(6) &4.66(2)&6.50(1)    \\
  0.85     &    1.810(1)      &2.655(2) & 3.72(1)&5.18(2) &7.73(5)   \\
  0.90     &     1.883(1)    &2.821(5) &4.020(5) &5.79(2) &9.21(1)    \\
  0.95     &     1.959(2)     &2.989(7) &4.32(1) &6.45(2)&10.86(3)     \\
  1.00     &     2.038(1)    &3.155(5) &4.69(3) &7.35(5) &13.94(3)     \\
\end {tabular}
\end{ruledtabular}
\end{table*}

\begin{table*}
\caption{Coordinates of  points in the fluid-fluid coexistence curve  as computed with GEMC simulations for the cases $q=0.4$ and $q=0.5$, both with $\Delta=0.3$. The superscripts I and II refer to values corresponding to the two coexisting phases rich in big and small disks, respectively. The error on the last significant figure is enclosed between parentheses.}
\label{Demix_q04_D03}
\begin{ruledtabular}
\begin{tabular}{ddddddd}
 \multicolumn{1}{c}{$\qquad x_1^{\text{I}}$}&\multicolumn{1}{c}{$\qquad\rho^{\text{I}}\sigma_1^2$}&\multicolumn{1}{c}{$\qquad x_1^{\text{II}}$}&\multicolumn{1}{c}{$\qquad\rho^{\text{II}}\sigma_1^2$}&\multicolumn{1}{c}{$\qquad\beta p\sigma_1^2$}&\multicolumn{1}{c}{$\qquad\beta \mu_1$}&\multicolumn{1}{c}{$\qquad\beta \mu_2$}\\
 \hline
 \multicolumn{7}{c}{$q=0.4$}\\
0.9856(4)&0.865(1)&0.187(1)&2.01(1)&8.26(3) &9.91(1) &2.12(1)\\
0.958(1)&0.842(2)&0.207(3)&1.82(3)&6.92(3) &8.38(1) &1.674(4)\\
0.939(3)&0.841(3)&0.260(3)&1.64(3)&6.59(5) &7.928(4) &1.543(4)\\
 \hline
 \multicolumn{7}{c}{$q=0.5$}\\
0.961(2)&0.794(3)&0.18(1)&1.52(5)&5.54(6) &7.08(2) &2.231(5)\\
0.909(9)&0.79(1)&0.26(2)&1.32(9)&5.04(8) &6.40(2) &1.982(9)\\
\end{tabular}
\end{ruledtabular}
\end{table*}


\end{document}